\documentclass[english]{iopart}

\usepackage{graphicx}
\usepackage{url}
\usepackage{iopams} 

\expandafter\let\csname equation*\endcsname\relax
\expandafter\let\csname endequation*\endcsname\relax

\usepackage{amssymb}
\usepackage[latin1]{inputenc}
\usepackage{amsmath}
\usepackage{epsfig}

\newcounter{fig}
\begin{document}

\title[Selected non-holonomic functions]
{\Large Selected non-holonomic functions in lattice statistical
 mechanics and enumerative combinatorics.}

\vskip .3cm

\author{ S. Boukraa$^\ddag$, 
J-M. Maillard$^\dag$}
\address{$^\ddag$  \ LPTHIRM and IAESB,
 Universit\'e de Blida, Algeria
}
\address{$^\dag$ LPTMC, UMR 7600 CNRS, 
Universit\'e de Paris 6, Sorbonne Universit\'es, 
Tour 23, 5\`eme \'etage, case 121, 
 4 Place Jussieu, 75252 Paris Cedex 05, France} 

\vskip .2cm 

Email: bkrsalah@yahoo.com   and   maillard@lptmc.jussieu.fr

\vskip .2cm 

\vskip .3cm 

{\em Dedicated to A. J. Guttmann, for his 70th birthday.}

\vskip .3cm 

\begin{abstract}

We recall that the full susceptibility 
series of the Ising model, 
 modulo powers of the prime $\, 2$, reduce to algebraic functions.
We also recall the non-linear 
polynomial differential equation obtained by Tutte for the generating 
function of the $\, q$-coloured rooted triangulations by vertices, which 
is known to have algebraic solutions for all the numbers of the 
form $\, 2 \,+2\, \cos(j\pi/n)$, the holonomic status of $\, q=\, 4$ being unclear.
We focus on the analysis of the $\, q=\, 4$ case, showing that
the corresponding series is quite certainly non-holonomic. Along the line of 
a previous work on the  susceptibility of the Ising model, we consider 
this $\, q=\, 4$ series modulo the first eight primes $\, 2$, $\, 3$, ... $\, 19$,
and show that this (probably non-holonomic) function reduces,  modulo these
primes, to algebraic functions. We conjecture that this probably non-holonomic
function reduces to algebraic functions modulo (almost) every prime, or power 
of prime numbers. This raises the question to see whether such remarkable 
non-holonomic functions can be seen as ratio of diagonals of rational functions,
or even algebraic functions of diagonals of rational functions. 

\end{abstract}


\vskip .1cm 

\noindent {\bf AMS Classification scheme numbers}: 03D05, 11Yxx, 33Cxx,  34Lxx,  34Mxx, 34M55,  39-04, 68Q70   

\vskip .2cm


\vskip .3cm

\newpage

\section{Introduction}
\label{introduction}

Our aim in this paper is to study the reduction modulo primes, or power of primes,
of certain differentially algebraic power series $\, F(x)\,  =\, \,  \sum c_n x^n$,
with integer coefficients of interest in physics.
Let us first recall that a power series $\, F(x)$ is called an algebraic series 
if it satisfies a polynomial relation
\begin{eqnarray}
\label{1alg}
P(x,\, F(x))\,  =\,\,   0, 
\end{eqnarray}
that a holonomic series satisfies a finite order linear differential equation
(here $\, P_i(x)$ denotes polynomials with integer coefficients, $\,  F^{(i)}(x)$
denotes the $i$-th derivative of $\, F(x)$)
 \begin{eqnarray}
\label{1hol}
\sum_{i=0}^k P_i(x) \cdot \, F^{(i)}(x) \, =\, \,  0.
\end{eqnarray}
The series $\, F(x)$ is called 
a differentially algebraic series if there exists a polynomial  $\, P$ 
such that $\, F(x)$ satisfies a polynomial differential equation 
\begin{eqnarray}
\label{1dalg}
P(x,\, F(x),\, F'(x),...,\,  F^{(k)}(x))\,  =\,\,   0.
\end{eqnarray}
A series is said to be non-holonomic if it is not solution of a linear
differential equation like (\ref{1hol}).
We will say that a series is an algebraic function modulo a prime if
there is a polynomial $\, P$
such that the series satifies equation (\ref{1alg}) modulo that prime.

In a previous paper~\cite{Auto} we have shown
that the full susceptibility of the Ising model,
which  is a {\em non-holonomic} 
function~\cite{bo-gu-ha-je-ma-ni-ze-08,bernie2010},  actually 
reduces to 
{\em algebraic functions modulo any powers of the prime $\, 2$}. 

Modulo $\, 2^r$, one cannot distinguish the full susceptibility
from some simple diagonals of rational functions~\cite{Auto}
which reduce to {\em algebraic functions} modulo $\, 2^r$. Modulo 
$\, 2^r$ these results can, in fact, be seen as being a consequence 
of the fact that, in the decomposition
of the full susceptibility in an infinite sum of $\, n$-fold 
$\tilde{\chi}^{(n)}$ integrals~\cite{wu-mc-tr-ba-76}, these 
$\tilde{\chi}^{(n)}$  are actually series with 
 integer coefficients, {\em with an overall $\, 2^n$ factor}. This 
may yield to a prejudice 
that these remarkable reductions to algebraic functions could only
take place modulo powers of the prime $\, 2$.

It is not clear if such a reduction of the full susceptibility 
to algebraic functions also takes place for 
{\em other primes or powers of primes}. At 
the present moment, the high or low temperature series of the 
full susceptibility modulo, for instance, prime $\, 3$, are not 
long enough to confirm, or discard the fact that the associated 
series could actually correspond 
to an algebraic function modulo $\, 3$. 

These exact results  shed new light on
 this iconic function in physics. 
They provide a strong incentive to {\em systematically 
study other non-holonomic series modulo primes} 
(or powers of primes), in theoretical physics. It 
is very important to see whether this 
is an exceptional result,  or 
the first example of a large set of selected 
non-holonomic functions in theoretical physics.

Remarkably long low-temperature and high-temperature 
series expansions~\cite{Iwan}, with respectively 
$\, 2042$ and $\, 2043$, coefficients have been obtained  for the 
susceptibility of the square Ising model using an iterative 
algorithm~\cite{Orrick}, the {\em polynomial growth} of that 
algorithm~\cite{Orrick} being a consequence of a 
{\em discrete Painlev\'e quadratic recursion}~\cite{Perk,Perk2,Perk3}.
Sometimes such algorithms with  polynomial growth are called ``integrable'' 
algorithms.  At the present moment the full susceptibility of the Ising 
model has only this ``algorithmic integrability'': 
{\em no non-linear differential equation}, or 
even functional equation~\cite{FuncEqua}, are 
known for that very important {\em non-holonomic} function in physics. 

Our aim in the following is to study other {\em non-holonomic} physical 
series {\em modulo primes}, or powers of primes. No non-linear differential 
equations are known for non-holonomic functions in lattice statistical 
mechanics, however, this is not the case in an almost undistinguishable 
domain of mathematical physics, namely 
enumerative combinatorics. In that respect, we must recall Tutte's study of 
triangulations equipped with a proper colouring~\cite{Tutte1,Tutte2,Tutte3},
his work culminating in 1982, when he proved that the series $\, H(w)$ 
counting $\,q$-coloured rooted triangulations by vertices satisfies a 
{\em non-linear polynomial differential equation}~\cite{Tutte5,Tutte6}:
\begin{eqnarray}
\label{Tutte}
\hspace{-0.95in}&&   \quad  \quad 
2 \, q^2 \cdot \, (1\, -q)  \cdot \, w \, 
+ \, \Bigl(q\, w \, + 10 \, H(w) \,
 -6 \, w \, { {d H(w)} \over {dw}} \Bigr)
 \cdot \,   { {d^2 H(w)} \over {dw}^2} 
  \\
\hspace{-0.95in}&&   \quad  \quad  \qquad   \qquad  
+\, q \cdot \, (4\, -q)  \cdot \, \Bigl( 20 \, H(w) \,
 - \,18 \, w \, { {d H(w)} \over {dw}} \,
 + 9 \, w^2 \, { {d^2 H(w)} \over {dw^2}} \Bigr)
\, \,\,  =  \, \, \, \, 0. \nonumber  
\end{eqnarray}
This $\, q$-family of {\em non-linear polynomial differential equations}
has a large number of remarkable properties. For instance, the series 
 $\, H(w)$ reduces to {\em algebraic functions} for all the well-known 
Tutte-Beraha numbers, and in fact, for all the numbers of the 
form\footnote[1]{These selected algebraic values of $\, q$ 
have been underlined many times on the standard scalar Potts 
model on euclidean lattices (the critical exponents 
are rational numbers, ...). They are such that a 
group of birational symmetries of the model, which is generically an 
infinite discrete group, degenerates
 into a {\em finite group}~\cite{Maillard,Hyperbolic,TrianglePotts,Rammal}.}
 $\, q \, = \, \, 2 \, +2 \, \cos(j\, \pi/m)$. This remarkable result 
first appeared in~\cite{Odlyzko} and was really proved by 
O. Bernardi and M. Bousquet-M\'elou in~\cite{Bernardi}. The Tutte-Beraha numbers
accumulate\footnote[2]{To some extent the study of
these remarkable numbers was a strategy in order to make some progress on 
the four-colour problem.} at the integer value $\, q=4$.  
Interestingly, the status of the series $\, H(w)$, at the integer value $\, q=4$,
{\em remains unclear\,}: if it is not an algebraic function, is it a holonomic function 
or a non-holonomic function ? 

Other one-parameter dependent {\em non-linear polynomial differential equations} 
have been found in an enumerative combinatorics framework (see for 
instance~\cite{Bernardi,Bernardi2,Bousquetcat,Courtiel}). Curiously, 
few analysis have been performed on the remarkable  non-linear 
differential equation (\ref{Tutte}). For instance, one does not 
know if the non-linear differential equation 
(\ref{Tutte}) fits with some {\em Painlev\'e property}.

We will focus, in this paper, on the study of equation (\ref{Tutte}), 
because of its historical importance as the  first example of exact 
{\em non-linear differential equation} in enumerative combinatorics, 
and as a toy model for the study of the susceptibility
of the Ising model, and, more generally, for the emergence of similar 
{\em non-linear differential equations in lattice statistical mechanics}. 
More specifically, we will focus on the analysis of the series $\, H(w)$ 
{\em at the integer value $\, q=4$}. We will show that even if this series 
is quite certainly {\em non-holonomic}, it, however, has a quite 
remarkable property, totally reminiscent of what we found on the
susceptibility of the Ising model~\cite{Auto}: this 
(probably non-holonomic) function is such that it actually 
{\em reduces to algebraic functions modulo the first eight primes}: 
$\, 2$, $\, 3$, $\, 5$, ...  $\, 19$, as well as 
powers of these primes. It is tempting to conjecture that this
(probably non-holonomic) function 
{\em reduces to algebraic function modulo (almost) every prime} 
(or every power of prime). This would be compatible with 
the scenario~\cite{Auto} that this series could be 
a simple ratio of diagonals of rational functions, or,
more generally, an algebraic\footnote{Note that rational
functions of diagonals of rational functions can be reduced to simple
ratio of diagonals of rational functions.} function  
of diagonals of rational functions~\cite{Auto}. 
Such kind of result is clearly a strong incentive to perform similar
studies on other non-linear differential equations  emerging
in enumerative combinatorics~\cite{Bernardi,Bernardi2,Bousquetcat,Courtiel},
or to obtain longer series (modulo some small primes $\, p=3$, ...)
for the susceptibility of the Ising model, or to study 
systematically (in a first step) ratio of diagonals of rational functions.

\vskip .2cm 

\section{A few remarks on the solutions of Tutte's 
non-linear differential equation (\ref{Tutte}).}
\label{definition}

\vskip .1cm 

Let us consider the series  $\, H(w) \, = \, \sum \, h_n \, w^n$,
solution of equation (\ref{Tutte}) which counts the $\, q$-coloured 
rooted triangulations by vertices.
Its coefficients are the number $\, h_n$ of rooted triangulations 
with $\, n$ vertices. They satisfy a remarkably simple 
{\em quadratic recurrence relation}\footnote[1]{As O. Bernardi and
M. Bousquet-M\'elou wrote it in~\cite{Bernardi2}, ``to date 
this recursion remains entirely
mysterious and Tutte's tour de force has remained isolated''.}:
\begin{eqnarray}
\label{Tutterec}
\hspace{-0.95in}&&   \quad  \quad \quad 
q \cdot \, (n+1)(n+2) \cdot \, h_{n+2} \, \,\, \, = \, \,  \, \,\,
q  \cdot \, (q\,-4)  \cdot \, (3\, n\, -1) \, (3\, n\, -2) \cdot \, h_{n+1} 
\nonumber  \\
\hspace{-0.95in}&&   \quad  \quad  \qquad   \qquad \quad  
\, + \, 2 \, \sum_{i=1}^{n} \, \,  i \cdot \, (i+1) \cdot \, 
(3\, n\, -3\, i\, +1) \cdot \,  h_{i+1} \, h_{n-i+2}, 
\end{eqnarray}
with the initial conditions $\, h_{0}  \, = \, \,0$, 
$\, h_{1}  \, = \, \,0$,  $\, h_{2}  \, = \, \, q\, (q-1)$. The 
number of proper $\, q$-colourings of a triangle is  
$\, h_{3}  \, = \, \, q \cdot \, (q-1)\, (q-2)$. 

This series  $\, H(w)$ reads
\begin{eqnarray}
\label{firstterms}
\hspace{-0.95in}&&  \quad \quad \quad  \, \, \,
H(w) \, \, = \, \, \,  \, 
q \cdot \, ( q-1 ) \cdot \,  {w}^{2} \, \,  \, 
+ \, \, q \cdot \, ( q-1 ) \,  ( q-2 ) \cdot \, 
\sum_{n=3}^{\infty} \, \, P_n(q) \cdot \,  w^n, 
\end{eqnarray}
where the  first terms of the sum reads:
\begin{eqnarray}
\label{firsttermsread}
\hspace{-0.95in}&&   \quad \quad 
\sum_{n=3}^{\infty} \, \, P_n(q) \cdot \, w^n  
\, \, = \, \, \, \, \, 
 {w}^{3} \, \, \, 
+  \, ( 4\,q-9 ) \cdot \,  {w}^{4} \,\,\,  
 +3  \cdot \, ( 8\,{q}^{2}-37\,q+43 ) \cdot  \,  {w}^{5}
\nonumber \\
\hspace{-0.95in}&&   \quad \quad \quad  \quad \quad 
\, +\,   (176\,{q}^{3}-1245 \,{q}^{2}+2951\,q-2344 ) \cdot \,  {w}^{6}\, 
 \\
\hspace{-0.95in}&&   \quad \quad  \quad  \quad \quad 
\,+ \,  (1456\,{q}^{4} -13935\,{q}^{3} +50273\,{q}^{2}
 -81036\,q +49248) \cdot \,  {w}^{7}
\, \,\, \, + \, \cdots 
\nonumber 
\end{eqnarray}
Of course there are many other solutions. For instance, with 
other initial conditions, namely
$\, h_{0}  \, = \, \,0$, but $\, h_{1}  \, \ne \, \,0$, 
one deduces a {\em one-parameter family of solutions}:
\begin{eqnarray}
\label{oneparam}
\hspace{-0.95in}&&   \quad 
H(w) \, \,  \,= \, \, \, \,  \, h_{1} \cdot \, w \, \, \,\,
+ \, \, q \cdot \, {{U} \over {q \, +4\cdot \, h_{1} }} \cdot \, w^2 \, \,\,
\, + \, \, {{ q^2 \cdot \, U \, V } \over { (q \, +4\cdot \, h_{1})^3 }}
\cdot \, w^3 \,  \cdot \, h(z), 
\end{eqnarray}
where:
\begin{eqnarray}
\label{oneparam2}
\hspace{-0.95in}&&  \quad  
U \,\, = \, \, \, q \cdot \, (q\, -1) \,  \,  +(q -4) \cdot \, h_{1}, 
\quad  \quad  
V \,\, = \, \, \, q \cdot \, (q\, -2) \,  \,  +2 \cdot \, (q -4) \cdot \, h_{1}.
\end{eqnarray}
and:
\begin{eqnarray}
\label{oneparamhz}
\hspace{-0.95in}&&  \quad  
 h(z) \, \, = \, \, \,  \,  \, 
1 \,  \,  \, \,   \,
+  \, \, \,  
\Bigl(q \,\cdot \,  (4\,q\,-9) \, +9 \cdot \,(q \,-4) \cdot \, h_{1}\Bigr)
  \cdot \, z
\, \, \,   \, +  \Bigl(129 \cdot \, (q-4)^2 \cdot \, h_{1}^2 \, 
 \nonumber  \\
\hspace{-0.95in}&&   \, \,  \,  \, \, \quad \quad \, 
+3 \cdot \, q\cdot \, (37 \, q-86)\cdot \, (q-4)\cdot \, h_{1} \,  \, 
+3 \cdot \, q^2\cdot \, (8\, q^2 \, -37 \, q\, +43)\Bigr) \cdot z^2
 \, \,  \, \, \, + \, \,  \,\cdots 
\nonumber \\
\hspace{-0.95in}&& \qquad \quad \quad 
\hbox{with:} \qquad \qquad \quad 
 \quad  z \, = \, \, {{ q \,\,  w} \over { (q \, +4 \,h_{1})^2 }}.   
\end{eqnarray}
When $\, U$ or $\, V$ in (\ref{oneparam}) are equal to zero, this yields 
two polynomial solutions of equation (\ref{Tutte}), valid 
{\em for any value} of $\, q$:
\begin{eqnarray}
\label{polsol}
\hspace{-0.95in}&&  \quad   \quad      \quad    \quad  
-{\frac {q \cdot \, (q-1) }{q-4}} \cdot \, w , 
\quad  \,  \,\quad \quad 
- \,{\frac {q \cdot \, (q \, -2) }{2 \, (q-4)}} \cdot \, w 
\,\, \, \, -\, {{ q \cdot \, (q-4)} \over {2}} \cdot \,  {w}^{2}.
\end{eqnarray}
Let us remark that, in the $\,h_{1} \, = \, 0$ limit, the 
series (\ref{oneparam})  reduces, for any value of $\, q$, 
to the series (\ref{firstterms}).

\vskip .1cm 

\subsection{The $\, q\, = \, 4$ subcase.}
\label{subcase}

In the $\, q\, = \, 4$ subcase the previous series 
(\ref{firstterms}) becomes:
\begin{eqnarray}
\label{serq4}
\hspace{-0.95in}&&  \, \, \quad     
H(w) \, \, = \, \, \,  \, 
12\,{w}^{2}  \, \, +24\,{w}^{3} \, +168\,{w}^{4} \, +1656\,{w}^{5} 
\, +19296\,{w}^{6} \, +248832\,{w}^{7} \, 
\nonumber \\
\hspace{-0.95in}&& \qquad \quad \quad \quad 
 +3437424\,{w}^{8} \, 
\, +49923288\,{w}^{9} \, +753269856\,{w}^{10} 
\,\, \, \, + \, \, \cdots 
\end{eqnarray}
If one considers the solutions of equation (\ref{Tutte}) with the 
initial conditions $\, h_{0}  \, = \, \,0$ and $\, h_{1}  \, = \, \,0$, 
but one does not impose $\, h_{2}  \, = \, \, q \cdot \, (q-1)  \, = \, \, 12$, 
one finds a {\em one-parameter family} of solutions of 
equation (\ref{Tutte}), namely 
(here $A$ denotes the parameter of this one-parameter family):
\begin{eqnarray}
\label{serq4onepara}
\hspace{-0.95in}&&  \, \, \quad     \quad   \quad  \quad    \quad     
H_A(w) \, \, \, = \, \, \, \,\, 
 -w \, \, \, \, 
+ \, A^3 \cdot \Bigl( {{w} \over {A^2}} 
\, \, + \,   H \Bigl( {{w} \over {A^2}} \Bigr) \Bigr), 
\end{eqnarray}
where the function $\, H$, in (\ref{serq4onepara}),
 is the previous series (\ref{serq4}). 
This corresponds to a one-parameter group of symmetry of the non-linear
differential equation  (\ref{Tutte}). Let us introduce the function 
$\, F(w) \, = \, H(w) \, + \, w$. It is solution of the (quite simple ...)
 non-linear differential equation:
\begin{eqnarray}
\label{newnonlin}
\hspace{-0.95in}&&  \, \,  \quad   \quad  \quad    \quad     \quad   
\Bigl(  3\cdot w \cdot \, {{ d F(w)} \over {dw}} \, -5 \cdot \, F(w) \Bigr)
 \cdot \, {{ d^2 F(w)} \over {dw^2}} 
\,\, \, \, \, + \, 48 \cdot \, w 
 \, \, \,\, = \, \, \, \,\, 0,
\end{eqnarray}
which has, clearly, the scaling symmetry 
$\,  F(w) \, \rightarrow \, A^3 \cdot F(w/A^2))$. This suggests
to define a function $\, G(w)$ such that 
$\, F(w) \, = \, w^{3/2} \cdot \, G(w)$.
Introducing the homogeneous derivative 
\begin{eqnarray}
\label{homder}
\hspace{-0.95in}&&  \, \, \, \, \quad \quad \quad  \quad  \quad   
G_1(w) \, = \, \, w \cdot \, {{d G(w)} \over {d w}}, \quad \quad  \quad  
G_2(w) \, = \, \, w \cdot \, {{d G_1(w)} \over {d w}}, \quad 
\end{eqnarray}
one finds that the non-linear
differential equation  (\ref{Tutte}), for $\, q\, = \, 4$, takes 
the very simple {\em autonomous}\footnote{As can be seen on equation (\ref{homder}) this equation
has constant coefficients.} form:
\begin{eqnarray}
\label{homder}
\hspace{-0.95in}&&  \, \, \, \, \quad \quad 
 (G(w) \, -6 \, G_1(w))\cdot \, 
(3\, G(w) \, +8\, G_1(w) \, +4\, G_2(w)) \, \, \,  -3\cdot \, 2^7
\, \,\, \,  = \, \,  \, \, 0.
\end{eqnarray}
As far as the singular points are concerned,
this change of function suggests that the exponent $\, 3/2$ 
should play a selected role.

In order to get very long series, we consider 
Tutte's recurrence (\ref{Tutterec}) for $\, q\, = \, \, 4$. Using this  
recurrence we have been able to get 24000 
coefficients\footnote[9]{This is a 376 Megaoctets file.} 
of the series (\ref{serq4}). This series has a finite radius of convergence 
 $\, r \simeq \, 0.04965 \,\, ...$, the coefficients growing like 
$\, \lambda^N$ where  $\, \lambda \simeq \,  \, 20.1378 \,\, ...$ 

\vskip .1cm

We first tried to see if such  very long series could actually correspond 
to a holonomic function using the same kind of tools we have already used in our 
(quite extreme) studies of $\, n$-fold integrals of the Ising 
type~\cite{bernie2010,Khi6,High}. We seek for linear differential 
operators, annihilating the series (\ref{serq4}) given with 
 $\, N$ coefficients ($N \, = \, 10000, \, \cdots, \, 24000$), 
of order $\, Q$ in the homogeneous derivative 
$\, \theta \, = \, \, w \cdot \, d/dw$ and of degree $\, D$ for
 the polynomial coefficients in front of the $\, \theta^n$'s, where
the order, degree, and number of coefficients are related by
a simple relation\footnote[1]{This kind of relation corresponds to the
so-called ``ODE formula'' see, for instance, equation (26) in~\cite{High}.}:
\begin{eqnarray}
\label{ODEformula}
\hspace{-0.95in}&&  \, \, \, \, \quad \quad \quad  \quad  \quad \quad   
(Q \, +1) \cdot \, (D \, +1) \,  \, = \, \, \, N \, \, - \, 1500. 
\end{eqnarray}
For the series with $\, N=\, 24000$ coefficients we explored all the values of 
the order $\, Q$ and degree $\, D$ related~\cite{High} by the ``ODE formula''
(\ref{ODEformula}), and failed to find a linear differential operator 
annihilating (\ref{serq4}). This seems to 
exclude the possibility that the series (\ref{serq4}) could 
be a holonomic function. 

A diff-Pad\'e analysis\footnote[3]{We do thank S. Hassani for 
providing this diff-Pad\'e analysis.} of this (probably non-holonomic) series 
gives a first set of singular points with their corresponding exponents.
One gets the first set of singularities, namely one real singularity
$\, w_1 \, = \, 0.04965 \, ...$, and several complex singularities 
$ \, 0.202837 \, ... \, \pm \, i \cdot \, 0.0964358 \, ... $,
$ \,0.470420 \, ... \, \pm \, i \cdot \, 0.37727 \, ... $, 
$ \, 0.86028 \, ... \, \pm \, i \cdot \, 0.92557 \,... $, 
$ \, 1.3784 \, ... \, \pm \, i \cdot \, 1.82295 \, ... $, 
$ \, 1.8007 \, ... \, \pm \, i \cdot \, 0.48740 \, ... $, 
$ \, 2.029904 \, ... \, \pm \, i \cdot \, 3.150337 \, ... $, 
all of them with the exponent $\, 3/2$, the exponents at infinity 
being $ \, -1/3, \, -2/3, \, -4/3, \, -5/3, \, ...$
It is possible that performing such kind of linear differential
analysis of a  (probably non-holonomic) series with longer series, 
one could, with higher order linear differential operators, see the 
emergence of more and more singularities: this could be a way 
to convince oneself that this series is non-holonomic. What is 
the validity of such a 
{\em linear approach for a typical non-linear function} is 
an open question, which certainly requires quite extensive 
studies\footnote[5]{In the spirit of the calculations
we performed in~\cite{Renorm}.} {\em per se}. Let us rather 
perform, in the following, some simpler clear-cut arithmetic 
calculations on this quite large series.

\vskip .1cm 

\section{Reduction of the $\, q\, = \, 4$ series modulo primes.}
\label{reduc}

Recalling the results of a previous paper~\cite{Auto} where 
we have shown that the full susceptibility of the Ising model, 
which is a {\em non-holonomic} 
function~\cite{bo-gu-ha-je-ma-ni-ze-08,bernie2010},  
actually reduces to 
{\em algebraic functions modulo any powers of the prime $\, 2$}. 
It is tempting to see if the series (\ref{serq4}), for $\, q=4$, 
actually reduces to  {\em algebraic functions} modulo the first eight primes 
$\, 2$, $\, 3$, ... $\, 19$. 

Since we have developed some tools~\cite{Khi6,High}
to find the (Fuchsian) linear differential operator 
annihilating a given series,  
let us first try (before seeking directly for algebraic relations
on this series, see next section \ref{algmodprime}) 
to see if this series (\ref{serq4}), modulo the 
first eight primes, is solution of a linear differential operator. 

Since the coefficients of the series are all divisible by $\, 12$,
and the series starts with $\, w^2$, we consider, instead of the series 
(\ref{serq4}), this series divided by $\, 12 \, w^2$, modulo  
the first primes $\, 2$, $\, 3$, ... $\, 17$, and seek for 
{\em linear differential operators} annihilating these series modulo 
primes. It is only because we have a prejudice that this $\, q=4$ 
series is ``very special'' that we perform such calculations.

\vskip .2cm 

{\bf Caveat:} Since we are going to use 
our tools~\cite{Khi6,High,ze-bo-ha-ma-04,ze-bo-ha-ma-05c,Special}
to find (Fuchsian) linear differential operators modulo rather 
small primes (the first eight primes), one may be facing a problem 
we do not encounter with our previous studies~\cite{Khi6,High} performed 
with rather large primes ($2^{15} \, -19\, = \, \, 32749$, ... ). Modulo 
a prime $\, p$, {\em any} power series with {\em integer coefficients}
is solution of the linear differential operators
 $\,\, \theta^p \, - \, \theta$, where  $\, \theta$ 
denotes the homogeneous derivative $\, w \cdot d/dw$, 
or much more simply of the operator $\, d^p/dw^p$.
Actually the linear differential operator, $\theta^p \, - \, \theta$ acting on $\, w^n$,
 gives (Fermat's little theorem):
\begin{eqnarray}
\label{caveat}
\hspace{-0.95in}&&   \quad     \quad \quad  \quad   \quad   \quad 
   (\theta^p \, - \, \theta)(w^n) 
\, \, = \, \, \, \, n^p \, -\,  n \, \, = \, \, \, \,0 
\quad  \quad  \quad   \quad  \bmod  \,\, \, p.
\end{eqnarray}
This is typically the reason why, when one is not in characteristic zero,
the wording ``being holonomic'' should be prohibited\footnote[1]{Because 
of identity (\ref{caveat}) every series is ``holonomic modulo a prime $\, p$'': 
one must seek for linear differential operators, getting rid of these 
spurious linear differential operators (\ref{caveat}).}.
When one performs such linear differential operator
guessing, modulo rather small prime $\, p$, it is important when 
one gets a result, to check, systematically, that the order of 
the linear differential operator one obtains,
 is strictly smaller than $\, p$, 
in order to avoid being ``polluted'' by such ``spurious'' 
linear differential operators. 

\vskip .2cm

\subsection{Reduction of the $\, q\, = \, 4$ series modulo 
the first eight primes: the results}
\label{reduceight}

To take into account the fact that all the integer 
coefficients of (\ref{serq4}) are divisible by $\, q \cdot \, (q-1) \, = \, 12$
we will consider, instead of (\ref{serq4}), the series (\ref{serq4})
divided by $\, 12\, w^2$:
\begin{eqnarray}
\label{norm}
\hspace{-0.95in}&& \quad \quad 
 S(w) \,\, = \, \,  \, {{H(w)} \over {12\, w^2}}
 \,\,\, = \, \, \, \, \,
1\,+2\,{w}\,+14\,{w}^{2}\,+138\,{w}^{3}\,
 +1608\,{w}^{4}\,+20736\,{w}^{5}
 \nonumber \\
\hspace{-0.95in}&& \quad \quad  \quad  \quad 
+\, 286452\,{w}^{6}\,+4160274\,{w}^{7} \,
 +62772488\,{w}^{8} \, +976099152\, w^{9} 
\,\, \,+ \, \cdots
\end{eqnarray}
From the previous recurrence relation (\ref{Tutterec}) for $\, q=\, 4$
 we obtained $ \, 24001$ coefficients of this series. 

We {\em actually found linear differential operators} for
 the series  (\ref{norm}), modulo  
the first primes $\,p= 2$, $\, 3$, ... $\, 17$. 
We denote 
$\, L_p$ the linear differential operators annihilating, modulo 
the prime $\, p$, the  series  (\ref{serq4}) divided by $\, 12 \, w^2$.
In the spirit of previous linear differential operator 
guessing~\cite{bernie2010,Khi6,High}, we introduce the homogeneous 
derivative $\, \theta \, = \, \, w \cdot d/dw$. The 
linear differential operators $\, L_p$ read 
respectively\footnote[9]{Modulo $\, 2$
 the series  (\ref{serq4}), divided by $\, 12 \, w^2$, is 
just the constant $\, 1$: the $\, L_2$ operator is trivially 
$\, \theta$. Slight transformations of the series have to be
performed to get a non-trivial result (see equation (\ref{mod2}) 
in section \ref{algmodprime} below).}: 
\begin{eqnarray}
\label{LP}
\hspace{-0.95in}&&   \quad     \quad       \quad 
 L_3 \, = \, \,  2\, w \,\, +\theta   \, +(w+1) \cdot \, \theta^2, 
\nonumber \\
\hspace{-0.95in}&&  \quad  \quad     \quad 
 L_5 \, = \, \, 2\, w \,\, +(2+3\, w) \cdot \, \theta \,\, 
 +(w+2)\cdot \,\theta^2,
 \nonumber \\
\hspace{-0.95in}&& \quad  \quad    \quad  
L_7 \, = \, \,3\, w^3 \,\, +(4+w^3) \cdot \, \theta \,\,
  +(3\,{w}^{3}+3)\cdot \, \theta^3 \,\,
+(5+w^3)\cdot \, \theta^4, 
 \nonumber
\end{eqnarray}
\begin{eqnarray}
\label{LP2}
\hspace{-0.95in}&&   \quad    \quad      \quad  
L_{11} \, = \, \, 9\,{w}^{15}+5\,{w}^{10}+5\,{w}^{5}\, \, 
+(2\,{w}^{15}+6\,{w}^{10}+9\,{w}^{5}+6) \cdot \, \theta \, 
\nonumber \\
\hspace{-0.95in}&& \quad \quad   \quad  \quad  \quad    \quad       
+(2\,{w}^{15}+8\,{w}^{10}+7\,{w}^{5}+1) \cdot \, \theta^2 \,\,
 +(5\,{w}^{15}+7\,{w}^{10}+{w}^{5}) \cdot \, \theta^3 \, 
 \nonumber \\
\hspace{-0.95in}&& \quad \quad   \quad   \quad   \quad   \quad      
+(6+4\, w^5+w^{10}+2\, w^{15})\cdot \, \theta^4 
\,\, +(10\,{w}^{15}+9\,{w}^{10}+8\,{w}^{5}+10) \cdot \, \theta^5
 \nonumber \\
\hspace{-0.95in}&& \quad \quad   \quad      \quad  \quad   \quad    
+(8\,{w}^{15}+8\,{w}^{10}+5\,{w}^{5}+7) \cdot \, \theta^6 \, \,
+(5\,{w}^{15}+4\,{w}^{5}+6) \cdot \, \theta^7 
\nonumber \\
\hspace{-0.95in}&& \quad \quad   \quad   \quad    \quad  \quad     
+({w}^{15}+{w}^{5}+8)\cdot \, \theta^8,
\end{eqnarray}
and: 
\begin{eqnarray}
\label{LP}
\hspace{-0.95in}&&   \quad  \quad  \quad    \quad    
 L_{13} \, = \, \, \sum_{n=0}^{8} \, p_n(w) \cdot \theta^n, 
\qquad \quad   \quad      
L_{17} \, = \, \, \sum_{n=0}^{13} \, q_n(w) \cdot \theta^n, 
\end{eqnarray}
where the polynomials $\, p_n$ and $\, q_n$ read respectively:
\begin{eqnarray}
\label{LPpn}
\hspace{-0.95in}&&    
p_0(w) \, = \, \, \, 
9\,{w}^{30}+8\,{w}^{27}+10\,{w}^{24}+11\,{w}^{21}+11\,{w}^{18}
+5\,{w}^{15}+10\,{w}^{12}+8\,{w}^{9}+2\,{w}^{6}, 
\nonumber \\
\hspace{-0.95in}&& 
p_1(w) \, = \, \, \, 
11\,{w}^{30}+4\,{w}^{27}+7\,{w}^{24}+4\,{w}^{21}+7\,{w}^{18}
+12\,{w}^{15}+{w}^{12}+2\,{w}^{9}+2\,{w}^{6} 
\nonumber \\
\hspace{-0.95in}&& \quad  \quad \quad  \quad     
 +9\,{w}^{3}+11, 
\nonumber \\
\hspace{-0.95in}&& 
p_2(w) \, = \, \, \, 
3\,{w}^{30}+7\,{w}^{27}+12\,{w}^{24}+2\,{w}^{21}+9\,{w}^{15}
+7\,{w}^{12}+5\,{w}^{9}+9\,{w}^{6}+2,
 \nonumber 
\end{eqnarray}
\begin{eqnarray}
\label{LPpnff}
\hspace{-0.95in}&& 
p_3(w) \, = \, \, \, 
6\,{w}^{30}+10\,{w}^{27}+7\,{w}^{24}+12\,{w}^{21}+9\,{w}^{18}
+10\,{w}^{15}+4\,{w}^{12}
\nonumber \\
\hspace{-0.95in}&& \quad  \quad \quad  \quad
 +2\,{w}^{9}+2\,{w}^{3}+6,
  \nonumber\\
\hspace{-0.95in}&& 
p_4(w) \, = \, \, \,
{w}^{30}+{w}^{27}+6\,{w}^{24}+6\,{w}^{21}+5\,{w}^{18}+2\,{w}^{15}
+7\,{w}^{9}+9\,{w}^{6}+8\,{w}^{3}+1,
  \nonumber\\
\hspace{-0.95in}&& 
p_5(w) \, = \, \, \,
12\,{w}^{30}+9\,{w}^{27}+4\,{w}^{24}+5\,{w}^{21}
+10\,{w}^{15}+3\,{w}^{12}
+3\,{w}^{9}+9\,{w}^{6}+6\,{w}^{3},
  \nonumber\\
\hspace{-0.95in}&& 
p_6(w) \, = \, \, \,
12\,{w}^{30}+{w}^{27}+7\,{w}^{24}+2\,{w}^{21}
+3\,{w}^{18}+9\,{w}^{15}
+2\,{w}^{12}+2\,{w}^{9}+5\,{w}^{6}
\nonumber \\
\hspace{-0.95in}&& \quad  \quad \quad  \quad     
+3\,{w}^{3}+1,
  \nonumber\\
\hspace{-0.95in}&& 
p_7(w) \, = \, \, \,
10\,{w}^{30}+6\,{w}^{24}+5\,{w}^{18}+3\,{w}^{15}+9\,{w}^{12}
+9\,{w}^{9}+3\,{w}^{6}+9\,{w}^{3}+9,
  \nonumber\\
\hspace{-0.95in}&& 
p_8(w) \, = \, \, \,{w}^{30}+6\,{w}^{27}+2\,{w}^{24}+2\,{w}^{18}
+10\,{w}^{15}+11\,{w}^{12}+7\,{w}^{9}+2\,{w}^{6}+2\,{w}^{3}+9, 
\nonumber
\end{eqnarray}
$\qquad$ and: 
\begin{eqnarray}
\label{LPqn}
\hspace{-0.95in}&&   
q_0(w) \, = \, \, \, 
15\,{w}^{40}+13\,{w}^{36}+2\,{w}^{32}+15\,{w}^{28}+16\,{w}^{24}
+7\,{w}^{20}+{w}^{16}+7\,{w}^{12}, 
\nonumber \\
\hspace{-0.95in}&& 
q_1(w) \, = \, \, \, 
15\,{w}^{40}+5\,{w}^{36}+5\,{w}^{32}+4\,{w}^{28}+12\,{w}^{24}
+15\,{w}^{20} +11\,{w}^{16}+2\,{w}^{12}
\nonumber \\
\hspace{-0.95in}&& \quad  \quad \quad  \quad     
 +15\,{w}^{8}+16\,{w}^{4}+5,
\nonumber \\
\hspace{-0.95in}&& 
q_2(w) \, = \, \, \, 
13\,{w}^{40}+9\,{w}^{36}+6\,{w}^{32}+{w}^{28}+5\,{w}^{24}+4\,{w}^{20}
+10\,{w}^{16}+5\,{w}^{12}
\nonumber \\
\hspace{-0.95in}&& \quad  \quad \quad  \quad     
 +4\,{w}^{8}+14\,{w}^{4}+15,
 \nonumber \\
\hspace{-0.95in}&& 
q_3(w) \, = \, \, \, 
15\,{w}^{40}+10\,{w}^{36}+12\,{w}^{32}+2\,{w}^{28}+14\,{w}^{24}
+10\,{w}^{20}+15\,{w}^{16}+5\,{w}^{12}
\nonumber \\
\hspace{-0.95in}&& \quad  \quad \quad  \quad     
+13\,{w}^{8}+10\,{w}^{4}+6,
 \nonumber 
\end{eqnarray}
\begin{eqnarray}
\hspace{-0.95in}&& 
q_4(w) \, = \, \, \, 
15\,{w}^{40}+{w}^{36}+4\,{w}^{32}+8\,{w}^{28}+13\,{w}^{24}
+6\,{w}^{20}+2\,{w}^{16}+8\,{w}^{4}+5,
 \nonumber \\
\hspace{-0.95in}&& 
q_5(w) \, = \, \, \,   
 4\,{w}^{40}+5\,{w}^{36}+11\,{w}^{32}+16\,{w}^{28}+13\,{w}^{24}
+6\,{w}^{20}+16\,{w}^{16}+{w}^{12}
\nonumber \\
\hspace{-0.95in}&& \quad  \quad   \quad   \quad         
+14\,{w}^{8}+4\,{w}^{4}+16,
 \nonumber \\
\hspace{-0.95in}&& 
q_6(w) \, = \, \, \, 
6\,{w}^{40}+9\,{w}^{36}+6\,{w}^{32}+11\,{w}^{28}
+{w}^{24}+8\,{w}^{20}
+6\,{w}^{16}+7\,{w}^{12}+4\,{w}^{8}
\nonumber \\
\hspace{-0.95in}&& \quad  \quad \quad  \quad     
+14\,{w}^{4}+11,
 \nonumber 
\end{eqnarray}
\begin{eqnarray}
\label{LPter}
\hspace{-0.95in}&& 
q_7(w) \, = \, \, \, 
14\,{w}^{40}+5\,{w}^{36}+11\,{w}^{32}+7\,{w}^{24}+8\,{w}^{20}
+11\,{w}^{16}+8\,{w}^{12}+2\,{w}^{8}
\nonumber \\
\hspace{-0.95in}&& \qquad \quad  \quad 
+11\,{w}^{4}+10,
 \nonumber \\
\hspace{-0.95in}&& 
q_8(w) \, = \, \, \, 
12\,{w}^{40}+5\,{w}^{36}+3\,{w}^{32}+6\,{w}^{28}+15\,{w}^{24}
+13\,{w}^{20}+16\,{w}^{16}+5\,{w}^{12}
\nonumber \\
\hspace{-0.95in}&& \quad  \quad \quad  \quad     
 +5\,{w}^{8}+11\,{w}^{4}+6,
 \nonumber 
\end{eqnarray}
\begin{eqnarray}
\label{LPquart}
\hspace{-0.95in}&& 
q_9(w) \, = \, \, \, 
14\,{w}^{40}+15\,{w}^{36}+11\,{w}^{32}+4\,{w}^{28}+14\,{w}^{24}
+{w}^{20}+14\,{w}^{16}+12\,{w}^{12}
\nonumber \\
\hspace{-0.95in}&& \quad  \quad \quad  \quad     
+13\,{w}^{8}+{w}^{4}+2,
 \nonumber 
 \end{eqnarray}
\begin{eqnarray}
\label{LPquint}
\hspace{-0.95in}&& 
q_{10}(w) \, = \, \, \, 
15\,{w}^{40}+16\,{w}^{36}+13\,{w}^{32} +13\,{w}^{28} +4\,{w}^{24}
+5\,{w}^{20}+6\,{w}^{16}+2\,{w}^{12}
\nonumber \\
\hspace{-0.95in}&& \quad  \quad \quad  \quad 
+9\,{w}^{8}+7\,{w}^{4}+13,
 \nonumber \\
\hspace{-0.95in}&& 
q_{11}(w) \, = \, \, \,
5\,{w}^{40}+2\,{w}^{36}+9\,{w}^{32} +13\,{w}^{28} +2\,{w}^{24}
+16\,{w}^{20}+11\,{w}^{16}+9\,{w}^{12}
\nonumber \\
\hspace{-0.95in}&& \quad  \quad \quad  \quad     
 +2\,{w}^{8}+4,
 \nonumber \\
\hspace{-0.95in}&& 
q_{12}(w) \, = \, \, \, 
9\,{w}^{40}+15\,{w}^{36}+14\,{w}^{28}+14\,{w}^{24} +8\,{w}^{20}
+10\,{w}^{16}+8\,{w}^{12}+12\,{w}^{8}
\nonumber \\
\hspace{-0.95in}&& \quad  \quad \quad  \quad
+14\,{w}^{4}+1,
 \nonumber \\
\hspace{-0.95in}&& 
q_{13}(w) \, = \, \, \, 
{w}^{40}+9\,{w}^{36}+9\,{w}^{32}+12\,{w}^{28}+6\,{w}^{24}
+12\,{w}^{20}+7\,{w}^{16}+14\,{w}^{12}
\nonumber \\
\hspace{-0.95in}&& \quad  \quad \quad  \quad     
 +9\,{w}^{8} +9\,{w}^{4} +8.
 \nonumber
\end{eqnarray}
We tried to get the linear differential operator $\, L_{19}$ 
for $\, p\, = 19$, but the calculations were too much 
time consuming. We will come to this $\, p\, = 19$ case 
with another more direct approach (see 
section \ref{seeking} below). 

\vskip .1cm 

It is quite a surprise to find {\em linear} differential operators
on such a {\em typically non-linear, probably non-holonomic}, function.
However, keeping in mind the results on the susceptibility 
of the Ising model~\cite{Auto}, it is natural to ask if such 
results modulo various primes could
correspond to reductions of the 
(probably non-holonomic) series (\ref{serq4}) 
to {\em algebraic functions} modulo primes.  This amounts 
to revisiting the previous series
modulo primes, trying to see, directly, if they are algebraic 
functions modulo primes, seeking for a polynomial equation
satisfied by these series modulo primes. Such calculations are performed
in the next section. An alternative way amounts to calculating
the {\em $\, p$-curvature}~\cite{bo-bo-ha-ma-we-ze-09}
 of these linear differential operators 
known modulo the prime $\, p$: if these series are reductions 
of algebraic functions modulo primes, the 
$\, p$-curvature~\cite{bo-bo-ha-ma-we-ze-09} 
{\em has to be equal to zero}. 

Taking into account the fact that the primes, considered here, are 
small enough, one can actually calculate the $\, p$-curvature using
some modular\footnote[2]{For larger prime numbers, one cannot, 
in practice, calculate the $\, p$-curvature that way, and one must 
use totally different algorithms~\cite{Fastalgo}.} 
algorithm~\cite{Cluzeau,Cluzeau2}. One actually finds that all these
linear differential operators $\, L_p$, modulo the primes $\, p$,
{\em  have zero $\, p$-curvature}\footnote[1]{We thank J-A. Weil for 
providing this result using a modular algorithm.}. 

\vskip .2cm 

\section{Algebraic functions modulo primes.}
\label{algmodprime}

Let us show that these series, modulo various primes, are actually 
{\em algebraic functions modulo primes}, by finding directly the polynomial 
equations they satisfy. 

Let us introduce the following {\em lacunary functions}
 which will be used in the following:
\begin{eqnarray}
\label{followlac}
\hspace{-0.95in}&& \quad \quad \quad 
{\cal L}_2(w) = \sum_{i=0}^\infty w^{2^i},  
\quad \quad \, \,\,  {\cal L}_3(w)= \sum_{i=0}^\infty w^{3^i},
\quad \quad  \, \,\,  {\cal L}_6(w)= \sum_{i=0}^\infty w^{2 \cdot 3^i}.
\end{eqnarray}

Similarly to the calculations performed in~\cite{Auto}
on the susceptibility of the Ising model, 
it is straightforward to see that, modulo the prime $\, 2$, 
a slight modification of the series (\ref{norm})
becomes the lacunary series $\, {\cal L}_2(w)$ which is  well-known 
to satisfy a functional equation and an algebraic equation, namely
 $\,\, \, {\cal L}_2(w^2) \, = \, \, {\cal L}_2(w) \, - \, w
 \, = \, \,  {\cal L}_2(w)^2 \, \,\, \,\,\, \bmod \, \, 2$. 

Modulo 2, we obtain: 
\begin{eqnarray}
\label{mod2}
\hspace{-0.95in}&&   \quad   \quad   \quad   \quad   \quad   \, \, 
{{w} \over {2}} \cdot \, (S(w)-1) 
\,\, \,  +\,w \cdot \, ( w^2 \, + \, 1 )
\,\, \, \, = \, \, \,\, {\cal L}_2(w).
\end{eqnarray}
Performing similar calculations, {\em modulo powers of the prime $\, 2$},
one gets similar results showing that the series {\em reduces to 
algebraic functions modulo powers of the prime $\, 2$}.

For instance, modulo $\, 2^2$, the following expression of $\, S(w)$ 
reduces, again, to the previous lacunary series:
\begin{eqnarray}
\label{mod22} 
\hspace{-0.95in}&&      \,  \quad \quad  \quad  \quad \quad 
{{w} \over {2}} \cdot \, (S(w)-1)  
\,\,\,\,+ \, w  \cdot \, (2 \, w^6 + \, \, w^2 +\, 1 ) \, 
\,\, \,\, = \, \, \, \, {\cal L}_2(w).
\end{eqnarray}
Modulo $\, 2^3$, one has:
\begin{eqnarray}
\label{mod23} 
\hspace{-0.95in}&&      \quad  \quad  \quad \quad  \quad 
w \cdot \, (S(w)-1)
\,\, \, + \, w \cdot \, ( 4 \, w^6 \, +2 \, w^2 +\, 2  )
\,\, \, = \, \,\, \,\, 2 \cdot \, {\cal L}_2(w).  
\end{eqnarray}
Modulo $\, 2^4$, one verifies on the series of $\, 24001$ coefficients,
 the following relation 
\begin{eqnarray}
\label{mod24} 
\hspace{-0.95in}&&      \quad  \quad \quad \quad  \quad 
w \cdot \, (S(w)-1) \,\,\, \,  = \, \, \, \, \,
(2 \, +8\, w) \cdot \, {\cal L}_2(w) 
\nonumber \\
\hspace{-0.95in}&& \quad \quad \quad  \quad \quad  \quad  \quad \quad 
\, + \, w \cdot \, (8\,{w}^{14}\,+4\,{w}^{6}\,+8\,{w}^{3}\,
+6\,{w}^{2}\,+8\,{w}\,+14).
\end{eqnarray}
Modulo $\, 2^5$, one can verify the more involved relation\footnote[3]{One may 
be surprised to see the occurrence of ${\cal L}_2^2$
in equation (\ref{mod25}) if one has in mind the identity ${\cal L}_2 = w +{\cal L}_2^2$. 
Note that this identity holds modulo $2$
 and {\em not} modulo $2^5$.}:
\begin{eqnarray}
\label{mod25} 
\hspace{-0.95in}&&     \quad  \quad 
w \cdot \, (S(w)-1)  \,\,\,\,  = \, \, \, \, \,
24 \cdot \, {\cal L}_2(w)^2 \,\, 
 +  (16\, w^3 \, +24\, w +26) \cdot \, {\cal L}_2(w) 
 \\
\hspace{-0.95in}&& \quad  \quad  \quad  \quad \quad 
\, + \, w \cdot \, ( 8\,{w}^{30} +4\,{w}^{14}+2\,{w}^{6} +8\,{w}^{5}
 +8\,{w}^{4}+4\,{w}^{3} +3\,{w}^{2}+12\,{w}+3).
\nonumber
\end{eqnarray}

\vskip .1cm  

Let us, now, consider the same series modulo the prime $\, 3$. One 
immediately sees the emergence of the lacunary series $\, {\cal L}_3(w)$: 
\begin{eqnarray}
\label{mod23h} 
\hspace{-0.95in}&&    \quad \quad    \quad   \quad \quad     
{{w} \over {2}} \cdot \, (S(w)-1) 
\,\, \,\,  + \, w \cdot \, (2 \, w \, +\, 1) \,  
 \,\, \, \,  = \, \, \, \, {\cal L}_3(w)
 \quad \quad \quad \bmod \, \, 3.
\end{eqnarray}

\vskip .1cm 

This  new lacunary series $\, {\cal L}_3(w)$ satisfies, modulo $\, 3$, a simple 
functional equation, as well as a simple algebraic equation
 $ \, {\cal L}_3(w^3) \, =  \, {\cal L}_3(w)  \, \, - w \, \, = \, \,{\cal L}_3(w)^3$. 
The series is thus an {\em algebraic function modulo $\, 3$}.

Modulo other primes (or power of primes) this guessing by lacunary series 
(along the line of~\cite{Auto}) is no longer well-suited. 

\vskip .2cm 

\subsection{Seeking for algebraic relations modulo primes.}
\label{seeking}

\vskip .1cm 

A better approach to analyse these series is to seek, 
systematically, modulo a given prime $\, p$,
 for a {\em polynomial relation}:
 $\, P(w, \, S(w)) \, = \, \, 0 \quad \bmod \, \, p$. 
\vskip .1cm 

As a first example, using
equation (\ref{mod23h}), one can see that the series $S(w)$ 
satisfies, modulo $\, p \, = \, \, 3$, 
the polynomial relation:
\begin{eqnarray}
\label{Rel3}
\hspace{-0.95in}&& \,  \, \, \quad  \quad  \,\, \,\,
w^2 \cdot \,  S(w)^{3} \,\, +2 \, \, S(w) \,\,
 \, \,+ \, (1+2\,w +{w}^{2} +{w}^{5}) 
\, \, \,= \,\, \,\, 0   
\quad \, \quad  \,  \, \bmod \, \, 3.
\end{eqnarray}

Modulo powers of the prime $\, p \, = \, \, 3$, one also obtains
reductions to algebraic functions, but the calculations 
are slightly more involved\footnote[3]{As we are going to see below, 
see equation (\ref{polynrelLac3actually}).}.
For instance, modulo $\, p \, = \, \, 3^2$ the series reads:
\begin{eqnarray}
\label{normbis3square}
\hspace{-0.95in}&& \quad \quad 
    S(w) \,\, = \, \,  \,  \,  \, 
1 \,\,\,   +2\,w \,+5\,{w}^{2} \,+3\,{w}^{3} 
\,+6\,{w}^{4} \,+6\,{w}^{7} \,
+8\,{w}^{8} \,+3\,{w}^{9} \,+3\,{w}^{11}
\nonumber \\
\hspace{-0.95in}&& \quad   \quad  \quad \quad \quad 
\,+8\,{w}^{26} \,+3\,{w}^{27}
\, +3\,{w}^{29} +3\,{w}^{35} +8\,{w}^{80} +3\,{w}^{81}
+3\,{w}^{83} +3\,{w}^{89}
\nonumber \\
\hspace{-0.95in}&& \quad   \quad  \quad \quad \quad 
 \,  +3\,{w}^{107} \,\; +8\,{w}^{242}
\, \, \, + \, \, \, \cdots 
\end{eqnarray}

In fact the series (\ref{normbis3square}) can actually 
be understood from the previously introduced lacunary 
series.  
The series (\ref{normbis3square}) can in fact be seen to be equal,
modulo $\, 3^2$, to: 
\begin{eqnarray}
\label{newlacequalto}
\hspace{-0.95in}&&  
 \, \,  \quad \quad 
 {{1}\over{w}} \, 
 \left( {{3}\over{2}}   \, {\cal L}_3^2 \,\, \, + \, 8  \, {\cal L}_3 \,
\, \, + \, 3  \, {\cal L}_6 \right) 
\,\,\, \,
+2 \, (3\,{w}^{7}+3\,{w}^{4}+3\,{w}^{2}+w+1).
\end{eqnarray}

Note that these lacunary series satisfy (in characteristic zero) 
the functional equations 
\begin{eqnarray}
\label{funcequ}
\hspace{-0.95in}&&  \quad  \, \,  \, \, \,
{\cal L}_3(w^3)  \, -{\cal L}_3(w) \,+w \,  \, =  \,  \, \, 0,
\quad \quad  \, \, \,
{\cal L}_6(w^3)  \, -{\cal L}_6(w) \,+w^2 \,  \, =  \,  \, \, 0.
\end{eqnarray}
Therefore these lacunary series satisfy, modulo $\, 3$, 
the polynomial relations:
\begin{eqnarray}
\label{polynrel}
\hspace{-0.95in}&&   \, \, 
{\cal L}_3^3  \, -{\cal L}_3 \,+w \,  \, =  \,  \, \, 0 
\, \,  \, \, \quad \bmod \, \, 3,
\quad \quad  \, \, \,\,
{\cal L}_6^3  \, -{\cal L}_6 \,+w^2 \,  \, =  \,  \, \, 0
\, \,  \, \, \quad \bmod \, \, \, 3.
\end{eqnarray}

\vskip .1cm 

The lacunary function $\,  {\cal L}_3$ satisfies, modulo $\, 3^2$,
the slightly more involved polynomial relation: 
\begin{eqnarray}
\label{polynrelLac3}
\hspace{-0.95in}&&   \, \, \quad \quad 
w^2 \, +\, w \cdot \, {\cal L}_3 \, +7 \cdot \, {\cal L}_3^2
 \,\,+2 \,  w \cdot  {\cal L}_3^3 \,\,
+\, {\cal L}_3^4 \,\,  +\, {\cal L}_3^6
\,  \,\, =  \,  \, \, 0
 \quad \, \, \,\,  \,\, \,\,  \, \bmod \, \,  \, 3^2.
\end{eqnarray}
Similarly, the lacunary function $\,  {\cal L}_6$ 
satisfies, modulo $\, 3^2$, the  polynomial relation: 
\begin{eqnarray}
\label{polynrelLac9}
\hspace{-0.95in}&&   \, \, \quad \quad 
w^4 \, \, +\, w^2 \cdot \, {\cal L}_6 \, \, +7 \cdot \,  {\cal L}_6^2
 \, \, +2   \,  w^2 \cdot  {\cal L}_6^3 \, \,
+\,  {\cal L}_6^4 \,  \, +\,  {\cal L}_6^6
\,  \,\, =  \,  \, \, 0 
\quad \, \, \, \,\, \,\, \, \, \,  \bmod \, \, \, 3^2.
\end{eqnarray}

The elimination of $\, {\cal L}_3$ and $\, {\cal L}_6$ 
in (\ref{newlacequalto})
gives a polynomial\footnote[1]{This polynomial 
 can easily be obtained
performing resultants in Maple.} relation of degree $\, 36$ in $\, S(w)$ 
and of degree $\, 72$ in $\,w$:
\begin{eqnarray}
\label{polynrelLac3}
\hspace{-0.95in}&&  \quad \quad \quad  \quad 
P(w, \, S(w)) \, \, = \, \, \,  \,
\sum_{n=0}^{36} \,  P_n(w)  \cdot S(w)^n 
\, \, \, = \, \, \, 0 \quad \quad  \quad  \bmod \, \, 3^2. 
\end{eqnarray}
We will not give this polynomial here because it is a bit too 
large. What matters is that it exists.  Now that we have these 
two degrees ($\, 36$ in $\, S(w)$ and $\, 72$ in $\,w$) for
a first example of polynomial relation, one can revisit this example
trying to find, directly, simpler  polynomial relations of lower degree 
(especially in $\, S(w)$).
One actually finds the following polynomial relation 
of degree $\, 6$ in $\, S(w)$
and degree $\, 17$ in $\, w$:
\begin{eqnarray}
\label{polynrelLac3actually}
\hspace{-0.95in}&& \quad \quad \, \, \, 
{w}^{3} \cdot \, (8\,{w}^{17} +6\,{w}^{14} +3\,{w}^{13} +6\,{w}^{12}
 +6\,{w}^{11} +6\,{w}^{10} +5\,{w}^{8}+3\,{w}^{6}+{w}^{5} 
\nonumber \\
\hspace{-0.95in}&& \quad \quad \quad \quad \quad \, \, \, 
 +3\,{w}^{4} +3\,{w}^{3} +2\,{w}^{2} +6\,w+3) 
\nonumber \\
\hspace{-0.95in}&& \quad \quad \quad \, \, \, 
 + \, ( 5\,{w}^{15}+{w}^{12}+5\,{w}^{11}+{w}^{10}+5\,{w}^{9}
+5\,{w}^{8}+5\,{w}^{6}+5\,{w}^{5}+6\,{w}^{3}) \cdot\, S(w)
 \nonumber \\
\hspace{-0.95in}&& \quad \quad \quad \, \, \, 
 +4\,{w}^{5} \cdot \, (2\,{w}^{5}+2\,{w}^{2}+w+2) \cdot S(w)^2
 \\
\hspace{-0.95in}&& \quad \quad \quad \, \, \, 
+{w}^{7} \cdot \,  (w^{10}+2\,{w}^{7}+{w}^{6}+2\,{w}^{5}+{w}^{4}+{w}^{3}+w+1) 
  \cdot \, S(w)^3
\nonumber \\
\hspace{-0.95in}&& \quad \quad \quad \, \, \, 
+{w}^{7} \cdot \, (2\,{w}^{5}+2\,{w}^{2}+w+2)   \cdot\, S(w)^{4}
\nonumber \\
\hspace{-0.95in}&& \quad \quad \quad \, \, \, 
+3\,{w}^{7}  \cdot\, S(w)^{5}
+{w}^{9} \cdot \, (2\,{w}^{5}+2\,{w}^{2}+w+2)  \cdot\, S(w)^{6}
\, \, \,= \, \,\, \, 0 
\quad \quad  \, \,   \bmod \, \, 3^2. 
\nonumber
\end{eqnarray}

Because of the quite large size of these  polynomial relations we will not, 
in the following,  give these relations corresponding to the series modulo 
power of primes for the next primes. 

\vskip .1cm 

Modulo  $\, p \, = \, \, 5$, we obtained the polynomial relation: 
\begin{eqnarray}
\label{Rel5}
\hspace{-0.95in}&& \quad \quad   \quad   \quad \, \, \, \, \, \, 
w \cdot \, S(w)^2 \,\, + S(w) \,\, \, +2\,{w}^{2}+2\,w+4
\, \, \,= \,\,\, \, 0   
\quad  \quad \quad \bmod \, \, 5. 
\end{eqnarray}
 
\vskip .1cm 

Modulo $\, p \, = \, \, 7$, we obtained the polynomial relation 
\begin{eqnarray}
\label{Rel7}
\hspace{-0.95in}&& \,   \, \,   \, \,\, \,   
{w}^{4} \cdot \, S(w)^{4} \, \, \,
+ \, {w}^{2} \cdot \, (5\,w+1)  \cdot \, S(w)^{3} \,\, \,  \, 
+ \, w \cdot \, (6\,{w}^{2}+5\,w+2)  \cdot \, S(w)^{2} \,
\\ 
 \hspace{-0.95in}&& \quad \,  \quad  \, \,     \quad   \quad  \quad  
+ \, ({w}^{2}+2\,w+6) \cdot \,  S(w) \, \, \,  +2\,{w}^{2}+5\,w+1 
\, \,\, = \,\, \, \, 0   \quad  \quad \, \, \, \,   \bmod \, \, 7. 
\nonumber
\end{eqnarray}

\vskip .1cm 

\vskip .1cm 

Modulo $\, p \, = \, \, 11$, we obtained the polynomial relation 
\begin{eqnarray}
\label{Rel11}
\hspace{-0.95in}&& \quad \quad \quad \quad \quad \quad 
\sum_{n=0}^{10} \, \, p_n(w) \cdot S(w)^n \, \, \, = \, \, \, 0, 
\quad \quad \quad \qquad \hbox{where:}
\nonumber \\
\hspace{-0.95in}&&\quad \quad \quad 
 p_0(w)  \,\, = \, \,\, 
{w}^{9}+4\,{w}^{8}+2\,{w}^{7}+9\,{w}^{6}+2\,{w}^{5}+{w}^{4}
+8\,{w}^{3}+8\,{w}^{2}+3\,w+3,
\nonumber \\
\hspace{-0.95in}&&\quad \quad \quad 
p_1(w)  \,\, = \, \,\, 
8\,{w}^{9}+8\,{w}^{8}+6\,{w}^{7}+7\,{w}^{6}+2\,{w}^{4}
+10\,{w}^{3}+4\,{w}^{2}+9\,w+8,
\nonumber \\
\hspace{-0.95in}&&\quad \quad \quad 
p_2(w)  \,\, = \, \,\, w \cdot \, 
(4\,{w}^{9}+{w}^{8}+2\,{w}^{7}+3\,{w}^{5}+7\,{w}^{4}
+4\,{w}^{3}+3\,{w}^{2}+9\,w+5), 
\nonumber \\
\hspace{-0.95in}&&\quad \quad \quad 
p_3(w)  \,\, = \, \,\, 
{w}^{2} \cdot \, (8\,{w}^{8}+10\,{w}^{7}+2\,{w}^{6}+{w}^{5}
+{w}^{4}+10\,{w}^{3}+4\,{w}^{2}+3\,w+5),
\nonumber \\
\hspace{-0.95in}&&\quad \quad \quad 
p_4(w)  \,\, = \, \,\, 
{w}^{3} \cdot \, (2\,{w}^{8}+2\,{w}^{7}+3\,{w}^{6}+2\,{w}^{5}
+{w}^{4}+8\,{w}^{3}+3\,{w}^{2}+8), 
\nonumber \\
\hspace{-0.95in}&&\quad \quad \quad 
p_5(w)  \,\, = \, \,\,
{w}^{4} \cdot \, (3\,{w}^{7}+9\,{w}^{6}+8\,{w}^{5}+5\,{w}^{4}+10\,{w}^{2}+6), 
\nonumber \\
\hspace{-0.95in}&&\quad \quad \quad 
p_6(w)  \,\, = \, \,\,
{w}^{7} \cdot \, (6\,{w}^{5}+10\,{w}^{4}+{w}^{3}+9\,{w}^{2}+9), 
\\
\hspace{-0.95in}&&\quad \quad \quad 
p_7(w)  \,\, = \, \,\, 2\,{w}^{10} \cdot \, (3\,{w}^{2}+5\,w+3), 
\nonumber \\
\hspace{-0.95in}&&\quad \quad \quad 
p_8(w)  \,\, = \, \,\, {w}^{12} \cdot \, (9\,w+1),  \quad \quad \,
p_9(w)  \,\, = \, \,\,  10\,{w}^{13}, \quad \quad \,
 p_{10}(w)  \,\, = \, \,\,  w^{14}. 
\nonumber 
\end{eqnarray}
One verifies that this polynomial equation is actually satisfied 
with our series of $\, 24001$ coefficients modulo $\, p\, = \, 11$.

\vskip .1cm

\vskip .1cm 

Modulo $\, p \, = \, \, 13$, we obtained the polynomial relation 
\begin{eqnarray}
\label{Rel13}
\hspace{-0.95in}&& \quad \quad \quad \quad \quad \quad 
\sum_{n=0}^{14} \, \, q_n(w) \cdot S(w)^n \, \, = \, \, \, 0, 
\quad \quad \qquad \, \, \hbox{where:}
\nonumber \\
\hspace{-0.95in}&&\quad \quad 
 q_0(w)  \,\, = \, \,\, 11\,{w}^{14}+6\,{w}^{13}+9\,{w}^{12}
+2\,{w}^{11}+6\,{w}^{10}+9\,{w}^{8} +10\,{w}^{7}+4\,{w}^{6}
\nonumber \\
\hspace{-0.95in}&&\quad \quad \quad  \quad \quad
 +4\,{w}^{5}+11\,{w}^{4}+11\,{w}^{3}+5\,{w}^{2}+10\,w+1, 
\nonumber \\
\hspace{-0.95in}&&\quad \quad 
 q_1(w)  \,\, = \, \,\, {w}^{14}+3\,{w}^{13}+7\,{w}^{12}
 +11\,{w}^{11}+3\,{w}^{10} +4\,{w}^{9}+8\,{w}^{8}+{w}^{7} +7\,{w}^{6}
\nonumber \\
\hspace{-0.95in}&&\quad \quad \quad  \quad \quad
 +5\,{w}^{5}+6\,{w}^{4}+5\,{w}^{3}+9\,w +12,
\nonumber \\
\hspace{-0.95in}&&\quad \quad 
 q_2(w)  \,\, = \, \,\, w \cdot \, 
(6\,{w}^{14}+2\,{w}^{13}+2\,{w}^{12}+11\,{w}^{8}+11\,{w}^{7}
+10\,{w}^{6} +{w}^{4}
\nonumber \\
\hspace{-0.95in}&&\quad \quad \quad \quad \quad 
+7\,{w}^{3}+11\,{w}^{2}+6\,w+9),
\nonumber \\
\hspace{-0.95in}&&\quad \quad 
 q_3(w)  \,\, = \, \,\, {w}^{2} \cdot \, 
(3\,{w}^{13}+6\,{w}^{12}+11\,{w}^{11}+6\,{w}^{10}+11\,{w}^{9}
+5\,{w}^{8} +5\,{w}^{7}+5\,{w}^{6}
\nonumber \\
\hspace{-0.95in}&&\quad \quad \quad \quad \quad 
+5\,{w}^{5}+4\,{w}^{4}+8\,{w}^{3}+9\,{w}^{2}+9\,w+1), 
\nonumber \\
\hspace{-0.95in}&&\quad \quad 
 q_4(w)  \,\, = \, \,\, {w}^{3} \cdot \, 
(9\,{w}^{13}+2\,{w}^{12}+9\,{w}^{11}+6\,{w}^{10}+10\,{w}^{8}
+12\,{w}^{7} +10\,{w}^{6}
\nonumber \\
\hspace{-0.95in}&&\quad \quad \quad \quad \quad 
+10\,{w}^{5} +7\,{w}^{4}+7\,{w}^{3}+5\,{w}^{2}+w+9), 
\nonumber 
\end{eqnarray}
\begin{eqnarray}
\label{Rel13bis}
\hspace{-0.95in}&&\quad \quad 
 q_5(w)  \,\, = \, \,\, {w}^{4} \cdot \, 
 (7\,{w}^{12}+11\,{w}^{11}+9\,{w}^{10}+4\,{w}^{9}+5\,{w}^{8}
+12\,{w}^{7} +7\,{w}^{6} +5\,{w}^{5}
\nonumber \\
\hspace{-0.95in}&&\quad \quad \quad \quad \quad 
+7\,{w}^{4}+5\,{w}^{3}+9\,{w}^{2}+12),
\nonumber \\
\hspace{-0.95in}&&\quad \quad 
 q_6(w)  \,\, = \, \,\, {w}^{5} \cdot \, 
({w}^{12}+{w}^{11}+12\,{w}^{10}+7\,{w}^{9} +4\,{w}^{8}
+3\,{w}^{7} +8\,{w}^{6}+4\,{w}^{5}
\nonumber \\
\hspace{-0.95in}&&\quad \quad \quad \quad \quad 
+5\,{w}^{4} +10\,{w}^{3}+2\,{w}^{2} +11), 
\nonumber 
\end{eqnarray}
\begin{eqnarray}
\label{Rel13bis2}
\hspace{-0.95in}&&\quad \quad 
 q_7(w)  \,\, = \, \,\, {w}^{8} \cdot \, 
(9\,{w}^{9}+8\,{w}^{8}+{w}^{7}+10\,{w}^{6}+2\,{w}^{5}
+6\,{w}^{4}+10\,{w}^{3}+12\,{w}^{2}+5),
\nonumber \\
\hspace{-0.95in}&&\quad \quad 
 q_8(w)  \,\, = \, \,\, {w}^{9} \cdot \, 
(7\,{w}^{9}+10\,{w}^{8}+{w}^{7}+2\,{w}^{6}+9\,{w}^{5}
+6\,{w}^{4}+2\,{w}^{3}+2\,{w}^{2}+1), 
\nonumber \\
\hspace{-0.95in}&&\quad \quad 
 q_9(w)  \,\, = \, \,\, {w}^{12} \cdot \, (w+1) \cdot \, 
 (5\,{w}^{5}-4\,{w}^{4}+4\,{w}^{3}+7\,{w}^{2}-w+1), 
\nonumber \\
\hspace{-0.95in}&&\quad \quad 
 q_{10}(w)  \,\, = \, \,\, 
{w}^{13} \cdot \, (6\,{w}^{6}+{w}^{5}+8\,{w}^{3}+6\,{w}^{2}+9), 
\, \,\quad \, \, \, 
q_{11}(w)  \,\, = \, \,\, {w}^{16} \cdot \, ({w}^{3}+6), 
\nonumber \\
\hspace{-0.95in}&&\quad \quad 
 q_{12}(w)  \,\, = \, \,\, {w}^{17} \cdot \, ({w}^{3}+6), 
\, \, \quad \quad \,\,
 q_{13}(w)  \,\, = \, \,\, 12\,{w}^{20}, \quad \, \,\quad \,\,
 q_{14}(w)  \,\, = \, \,\, {w}^{21}. 
\nonumber 
\end{eqnarray}
One verifies that this polynomial equation is actually satisfied 
with our series of $\, 24001$ coefficients modulo $\, p\, = \, 13$.

\vskip .1cm 

Modulo $\, p \, = \, \, 17$, we obtained the polynomial relation 
\begin{eqnarray}
\label{Rel17}
\hspace{-0.95in}&& \quad \quad \quad \quad \quad \quad  \quad  \quad 
\sum_{n=0}^{24} \, \, r_n(w) \cdot S(w)^n \, \, \, = \, \, \, 0, 
\end{eqnarray}
where the polynomials $\, r_n(w)$ are given in \ref{pol17}.
One verifies that this polynomial equation is actually satisfied with 
our $\, 24001$ coefficients series modulo $\, p\, = \, 17$.

\vskip .1cm 

Modulo $\, p \, = \, \, 19$, we obtained the polynomial relation 
\begin{eqnarray}
\label{Rel19}
\hspace{-0.95in}&& \quad \quad \quad \quad \quad \quad  \quad \quad  
\sum_{n=0}^{30} \, \, s_n(w) \cdot S(w)^n \, \, \, = \, \, \, 0, 
\end{eqnarray}
where the polynomials $\, s_n(w)$ are given in  \ref{pol19}.
One verifies that this polynomial equation is actually satisfied with 
 $\, 23756$ coefficients of our series.

After this accumulation of 
algebraic results, it seems reasonable to conjecture that the series 
(\ref{serq4}), or equivalently (\ref{norm}), {\em reduces to 
algebraic functions modulo every prime} (and probably modulo power 
of primes, but it is much more difficult
to confirm this statement modulo power of primes).

\vskip .2cm 

\vskip .1cm 

{\bf Remark:} When one does not restrict to primes
the results have to be taken ``cum grano salis''.  For instance 
modulo $\, 6$, the series  modulo $\, 6$ reads: 
\begin{eqnarray}
\label{SeRel6cub}
\hspace{-0.95in}&&  \quad \quad  \, \, 
S(w) \, \,   = \, \,  \, \, 
1\, \, \,  +2\,w\, \,+2\,{w}^{2}\, +2\,{w}^{8}\, +2\,{w}^{26}\, +2\,{w}^{80}\, 
+2\,{w}^{242}\, +2\,{w}^{728}
\nonumber \\
\hspace{-0.95in}&& \quad \quad \quad \quad \quad  \quad \quad \quad 
 +2\,{w}^{2186}\, \,
 +2\,{w}^{6560}\, \,  \, \, + \, \, \, \cdots 
\end{eqnarray}
If one considers the expression
 $ \,\, w \cdot (1\, +S(w))/2  \,  -  \, w^2 \, $, 
one actually finds that it is nothing but 
the selected lacunary series $\, \sum \, w^{3^n} = {\cal L}_3 (w)$:
\begin{eqnarray}
\label{SeRel6cubS4cons}
\hspace{-0.95in}&&   \quad 
 {{w} \over {2}} \cdot \, \bigl( 1\, +S(w) \bigr) \, - \, w^2
 \, \,  \,  \,  =  \, \, \,
{\cal L}_3 (w) \, \, \, =
\\
\hspace{-0.95in}&&  \quad  \quad  \quad 
w \,+{w}^{3} 
\,+{w}^{9} \,+{w}^{27} \,+{w}^{81} \,+{w}^{243} 
\,+{w}^{729} \,+{w}^{2187} \,+{w}^{6561} +w^{19683}
 \, \,\, + \, \, \, \cdots 
\nonumber 
\end{eqnarray}
Following the ideas displayed in~\cite{Pacific}, one
can see that this series is not algebraic modulo $\, 6$.
This series $\, S(w)$ is algebraic modulo $\, 3$ 
(because $S(w^3)\, = \, \,  w \, +S(w)$ and  
$S(w)^3\, = \, \, S(w^3) \quad \bmod \, \, 3$), but it is not 
algebraic modulo $\, 2$. If it were algebraic modulo $\, 2$,
it would be\footnote[1]{There is a theorem by Cobham~\cite{Cobham} 
which says that if a series has only coefficients $\, +1$ it can 
be algebraic modulo two successive primes (here $\, 2$ and $\, 3$)
only if it is rational. Furthermore if a series is algebraic
 modulo two relatively prime numbers, 
namely a prime $\, p$ and also another prime $\, q$, 
it is algebraic modulo $\, p \cdot q$.} algebraic modulo $\, 6$.

\vskip .1cm
 
\vskip .1cm 

\section{Comparison with other reductions modulo primes.}
\label{compare}

This first example of reduction to algebraic functions modulo primes,
or power of primes, of (probably {\em non-holonomic}) functions, 
satisfying {\em non-linear differentiable equation}s, is unexpected
in a more general non-holonomic, non-linear framework.

In order to have some perspective on such kind of results, let us 
consider series with {\em integer coefficients},
 that are solutions of {\em linear} differential equations 
(holonomic). Let us consider  
{\em diagonals of rational 
functions}~\cite{Poorten,Denef,Adamczewski,Lairez,Lairez1,Rowland}, 
and also holonomic {\em globally bounded} $\, G$-series which 
are {\em not known} to be diagonals of rational functions~\cite{Short,Big}.

\vskip .1cm 

\subsection{Reductions modulo primes of holonomic functions: 
diagonals of rational functions and beyond}
\label{compare2}

\vskip .1cm 

Diagonals of  rational functions are known to reduce to algebraic functions 
modulo any prime~\cite{Short,Big} (or power of primes). Reductions 
modulo primes of diagonals of rational functions 
are, in general, quite easy and quick to perform. When
the order of the linear differential operator is not too large
one gets quite easily the algebraic functions corresponding to this 
reduction. One should note that diagonals of rational functions that are 
$\, _nF_{n-1}$ {\em hypergeometric series} are ``almost too simple''
(see \ref{appendixhyper}).
The reduction of hypergeometric series are, most of the time, very simple 
algebraic functions of the form $\, P(x)^{-1/N}$ (where $\, N$ is an integer
and $\, P(x)$ is a polynomial), which correspond to the truncation of the 
series expansion of the hypergeometric series modulo the prime $\, p$.
We sketch a few results of such reductions of hypergeometric series 
modulo primes in \ref{appendixhyper}.

\vskip .2cm

Along this  hypergeometric line it is worth recalling the hypergeometric
function 
$\, _3F_2([1/9, \, 4/9, \, 5/9],[1/3, \, 1], 3^6 \, x)$
introduced by 
G. Christol~\cite{Short,Big,Christol}, a few decades ago, 
to provide an example of holonomic $\, G$-series with {\em integer coefficients} 
that {\em may not be the diagonal of rational function}. After all 
these years, it is still an open question to see whether this 
function is, or is not, the diagonal of rational function. In such 
cases it is not guaranted\footnote[2]{The question to know
if globally bounded $\, D$-finite
formal power series (non-zero radius of convergence) are
globally automatic (their reduction modulo all
but finitely many primes $\, p$ is $\, p$-automatic), 
{\em remains an open question}: see Question and Remark 
page 385 of~\cite{Allouche}.}  
that the corresponding series modulo primes are algebraic 
functions (or that the series are ``automatic''~\cite{Auto}).

If one performs the same reductions modulo primes, one finds,  in contrast 
with the previous studies of reductions modulo primes of diagonals of 
rational functions, that it becomes 
{\em quite hard to see whether 
a series like $_3F_2([1/9,4/9,5/9],[1/3,1],3^6 x)$, modulo primes are algebraic functions} 
(they could be of the form $\, P(x)^{-1/N}$ where $\, N$ is an extremely 
large integer, see \ref{holoreduc}, or they could satisfy polynomial relations 
of the ``Frobenius'' type of large degree, see \ref{holoreduc}).
Probably different strategies ($p$-automatic approaches) 
should be considered to find these polynomial relations (if any). 

\vskip .1cm 

\vskip .1cm 

{\bf To sum up:} As far as the reduction of {\em holonomic functions} 
modulo primes is concerned, we seem to have the following situation: 
either the  holonomic function is actually the {\em diagonal 
of a rational function}~\cite{Short,Big}, the  reduction to 
algebraic function modulo primes is thus guaranted,
and one finds, very simply and quickly, these algebraic functions, 
or  the  holonomic function 
is {\em not} ``obviously'' the diagonal of the rational function,
 and getting these 
 algebraic functions can be very difficult (see \ref{holoreduc}).

\vskip .1cm 

This difficulty to find polynomial relations modulo rather small primes, 
for such a holonomic function (which is not obviously the diagonal of 
a rational function), has to be compared with the rather easy way 
we obtained, in section \ref{seeking},  polynomial relations
for a (probably {\em non-holonomic}) series solution of the $\, q=4$
{\em non-linear differential equation} (\ref{Tutte}).

\vskip .1cm 
\vskip .1cm 

{\bf Remark:}  Modulo a prime $\, p$ we  have  linear differential 
operators of two\footnote[1]{In fact three if one takes into account the
``spurious'' linear differential operators (\ref{caveat}).} different
 natures annihilating  a given diagonal 
of rational function: one has  linear differential operators of
 {\em nilpotent $\, p$-curvatures}~\cite{bo-bo-ha-ma-we-ze-09} (which 
are the reduction, modulo $\, p$, of the globally nilpotent 
 linear differential operators~\cite{bo-bo-ha-ma-we-ze-09}
 annihilating the series in characteristic zero), and 
one also has  linear differential operators of {\em zero $\,p$-curvatures},
corresponding to the fact that a diagonal of rational function reduces to algebraic 
functions modulo a prime $\, p$. For holonomic functions (in our case globally 
bounded~\cite{Short} $\, G$-series), 
the order of the  linear differential 
operator  (of nilpotent $\, p$-curvature) ``saturates''
with the order of the  linear differential operator in characteristic zero. 
In contrast, for {\em selected  non-holonomic functions, reducing 
to algebraic functions modulo primes},
one just has the second set of linear differential operators of zero $\,p$-curvature,
their order having  {\em no reason to have such an upper bound}. 
Increasing the value of the prime $ \, p$ in the  modular guessing of the 
linear differential operator could, thus, be a way to {\em disentangle 
between holonomic functions and selected non-holonomic functions} 
reducing to algebraic functions modulo primes. 

\vskip .1cm 

\vskip .1cm 

\subsection{Reductions modulo primes of other selected non-holonomic functions.}
\label{compare6}

One would like to accumulate more examples of reductions modulo primes of other 
selected {\em non-holonomic} functions. 
In an integrable lattice model perspective where the theory of elliptic 
curves plays so often a crucial role (as well as mirror
symmetries),  a quite natural candidate amounts to 
considering the ratio of two selected holonomic functions, namely the 
{\em ratio of two periods of an elliptic curve}~\cite{CalabiYauIsing,Singularities}.
Unfortunately, as can be seen in \ref{versus}, one cannot perform such a 
reduction because one of the two holonomic functions, in such a ratio, is 
{\em not globally bounded}~\cite{Short,Big}, which means that the series
cannot be recast into a series with integer coefficients: one cannot 
consider such series modulo primes\footnote[1]{An infinite number 
of primes occurs at the denominator of the successive coefficients 
of the series, preventing to consider such series modulo 
this infinite set of primes.}.

Therefore let us rather consider non-holonomic functions that are, 
{\em not only ratio of holonomic functions, but, in fact,  ratio
of diagonals of rational functions}. Let us consider, for instance,
 the ratio of two simple hypergeometric functions 
{\em that are diagonals of rational functions}~\cite{Short,Big}:
\begin{eqnarray}
\label{reduc7A}
\hspace{-0.75in}&& \quad \quad \quad \quad \quad \quad
R(x) \, \, = \, \,  \,  \, \, 
{{ _2F_1\Bigl([{{1} \over {3}}, \,{{1} \over {3}}], \, [1], \, \, 27 \, x \Bigr) 
} \over {
 _2F_1\Bigl([{{1} \over {2}}, \,{{1} \over {2}}], \, [1], \, \, 16 \, x \Bigr) }}.
\end{eqnarray}
This ratio {\em satisfies a non-linear differential equation} that can be
obtained from the two order-two linear differential equations satisfied
by these two  simple hypergeometric functions. We give this non-linear 
differential equation in \ref{Nonlin}. 

The series expansion of this ratio (\ref{reduc7A}) 
is a series with {\em integer coefficients}:
\begin{eqnarray}
\label{reduc7Aser}
\hspace{-0.95in}&& 
R(x) \, \, = \, \, \,  \, 
1 \, \,\,  -x \, \,  +4\,{x}^{2}+208\,{x}^{3}
 +5549\,{x}^{4} +133699\,{x}^{5} +3142224\,{x}^{6}
+73623828\,{x}^{7}
\nonumber \\
\hspace{-0.95in}&& \quad  \quad  \quad  \quad 
+1733029548\,{x}^{8} +41095725700\,{x}^{9}\, 
+982470703424\,{x}^{10}
 \, \, + \, \, \, \cdots 
\end{eqnarray}

These two  hypergeometric functions are
 diagonals of a rational function: their reductions modulo 
primes must be algebraic functions. 
For instance, modulo $\, p \, = \, 7$, it reads:
\begin{eqnarray}
\label{reduc7AB}
\hspace{-0.95in}&& \quad \quad \quad \, 
_2F_1\Bigl([{{1} \over {2}}, \,{{1} \over {2}}], \, [1], \, \, 16 \, x \Bigr) 
\, \, = \, \, \,  (1 \, +4\,{x}\,  +\,{x}^2 \, +{x}^3)^{-1/6}
\qquad  \,  \,  \bmod \, \,  7,
\\
\label{reduc7ABC}
\hspace{-0.95in}&& \quad \quad \quad \, 
_2F_1\Bigl([{{1} \over {3}}, \,{{1} \over {3}}], \, [1], \, \, 27 \, x \Bigr) 
\, \, = \, \, \,  (1 \, +\,3 \, {x}\,  +\, {x}^2)^{-1/6}
\qquad  \quad  \quad   \, \,  \bmod \, \,  7.
\end{eqnarray}

If one considers the {\em non-holonomic} series (\ref{reduc7Aser})  
corresponding to their ratio (\ref{reduc7A}),
it reduces modulo the prime $\, 7$, as it should,  to the ratio
of the (algebraic) reductions (\ref{reduc7AB}) and  (\ref{reduc7ABC}): 
\begin{eqnarray}
\label{reduc7ABff}
\hspace{-0.95in}&& \quad \quad \quad \quad \quad \quad \, \, \,\, \,
R(x) \, \, = \, \, \, 
\Bigl({{ 1 \, +4\,{x}\,  +\,{x}^2 \, +{x}^3
} \over {
 1 \, +\,3 \, {x}\,  +\, {x}^2}}\Bigr)^{1/6}  
 \quad   \quad    \, \, \,   \bmod \, \,  7.
\end{eqnarray}

\vskip .1cm 

The set of {\em non-holonomic series with integer coefficients,
reducing to algebraic functions modulo every prime} (or power of prime),
is clearly a very large set. 

\vskip .1cm 

\section{Conclusion}
\label{conclusion}
 
We have recalled that the full susceptibility series of the Ising model 
satisfies, modulo powers of the prime $\, 2$, exact algebraic
equations~\cite{Auto} which is a consequence of the fact
that, modulo $\, 2^r$, one cannot distinguish the full susceptibility
from some simple diagonals of rational functions which reduce to 
algebraic functions modulo $\, 2^r$. We also recalled the non-linear 
polynomial differential equation (\ref{Tutte}) 
obtained by Tutte for the generating 
function of the $\, q$-coloured rooted triangulations by vertices.

Along the line of a previous work~\cite{Auto} on the  susceptibility 
model, we considered this series, solution of (\ref{Tutte}), modulo 
the first eight primes $\, 2$, $\, 3$, ... $\, 19$,
and showed that this (probably non-holonomic) function actually
reduces, modulo these
primes, to algebraic functions. We conjecture that this
 {\em probably non-holonomic}
function {\em reduces to algebraic functions modulo (almost) every primes}, 
or power of primes, numbers. 

We believe that this result on the $\, q=4$ solution of
Tutte's non-linear differential equation (\ref{Tutte}) for the generating 
function of the $\, q$-coloured rooted triangulations by vertices, is not
an isolated curiosity, but corresponds to a first pedagogical example 
of a large class of remarkable non-holonomic functions in theoretical 
 physics (lattice statistical physics, enumerative combinatorics ...)
that reduce to algebraic functions modulo primes (and power of primes).
It is important to understand these remarkable non-holonomic functions:
are they ratio of holonomic functions (having in mind 
{\em ratio of diagonals of rational functions}), or 
more generally algebraic functions
of diagonals 
of rational functions~\cite{Auto}, do the non-linear differential equations
they satisfy have the {\em Painlev\'e property}\footnote[3]{One can actually show that
non-linear equation (\ref{Tutte})  {\em does not have the Painleve property}.
We thank A. Ramani and R. Conte for two different proofs
of this result.}, etc ... ? 

It is essential to build new tools, new algorithms to see whether a given
(large) series is solution of a {\em non-linear differential equation}, and, 
in particular, of a {\em polynomial differential equation}.
Too often Rubel's universal equation\footnote[1]{Rubel's non-linear 
differential equation~\cite{Rubel2} corresponds to a {\em homogeneous} 
polynomial differential equation such that any continuous function can be 
approximated, {\em on the real axis},  by a solution of this 
``universal'' equation. Other examples were obtained~\cite{Rubel} which 
correspond to the idea of piecewise polynomial approximation on the real axis. 
This kind of {\em real analysis} theorem {\em do not mean} that any function of a 
complex variable is ``almost'' solution of a non-linear differential equation
in the complex plane, which would mean that any 
``non-linear differential Pad\'e'' would be pointless.
} is recalled to discourage
any such ``{\em non-linear differential Pad\'e} '' search. 
It must be clear that this kind of ``non-linear differential Pad\'e'' analysis, 
should not be performed in the most general non-linear framework: it
must be performed with some assumptions, ansatz, corresponding to the problem
of theoretical physics one considers (Painlev\'e property 
assumption~\cite{Manin}, 
regular singularities assumptions, autonomous assumptions,
 see (\ref{homder}), 
non-linear differential equations associated with 
{\em Schwarzian derivatives}~\cite{CalabiYauIsing,Chakravarty,Chazy1,Chazy2}
 or 
{\em modular forms}~\cite{Harnad2,Harnad,Sebbar,Maier,Ramanujan,Ramanujan2},
  ...). 

It is crucial to build new tools, new algorithms to see whether a given
(large) series is a ratio of holonomic functions (having in mind 
ratio of diagonals of rational functions), or more generally 
algebraic functions of diagonals of rational functions.

Such kind of result is clearly a strong incentive
 to obtain longer series (modulo some small primes $\, p=3$, ...)
for the full susceptibility of the Ising model to see if the
susceptibility series reduces, for instance modulo $\, 3$, 
to an algebraic function.

 \vskip .1cm 

 \vskip .4cm 

 \vskip .1cm 

{\bf Acknowledgments:} 
We thank one referee for very detailed comments and suggestions.
We also thank A. Ramani and R. Conte for their help to clarify the
nature of equation (\ref{Tutte}).
One of us (JMM) would like to thank 
M. Bousquet-M\'elou  for fruitful discussions on Tutte's equation, 
and G. Christol for detailed discussions on diagonals of rational 
functions modulo power of primes and bringing to our attention
Cobham's theorem. He also thanks J-P. Allouche for stimulating 
``automatic'' discussions on diagonals of rational functions, and 
J-A. Weil for providing some 
$\, p$-curvature calculations. We thank A. Bostan for providing an 
algebraic result.
We thank S. Hassani  for providing a differential
Pad\'e analysis and for very helpful comments. 
This work has been performed without
 any support of the ANR, the ERC, the MAE or any PES of the CNRS. 

 \vskip .2cm 

 \vskip .5cm 

\appendix

\section{ Polynomial relations modulo $\, p \, = \, 17$ and $\, p \, = \, 19$}
\label{appendix}

Let us give the two polynomial relations satisfied by 
$\, S(w) \, = \,  \, H(w)/(12 \, w^2)$,  namely the series (\ref{norm}), 
modulo $\, p \, = \, 17$ and $\, p \, = \, 19$.

 \vskip .1cm 

\subsection{Polynomial relation for $\, p \, = \, 17$}
\label{pol17}

Modulo $\, p \, = \, \, 17$, we obtained the polynomial relation 
\begin{eqnarray}
\label{Rel17}
\hspace{-0.95in}&& \quad \quad \quad \quad \quad \quad 
\sum_{n=0}^{24} \, \, r_n(w) \cdot S(w)^n \, \, = \, \, \, 0, 
\qquad \quad \quad 
% \hbox{where:} 
\end{eqnarray}
where:
\begin{eqnarray}
\label{Rel17r}
\hspace{-0.95in}&&
 r_0(w)  \,\, = \, \,\,{w}^{27}+10\,{w}^{26}+9\,{w}^{25}
+8\,{w}^{24}+14\,{w}^{23}+12\,{w}^{22}
+{w}^{21}+7\,{w}^{20}+8\,{w}^{19}
\nonumber \\
\hspace{-0.95in}&&\quad \quad \quad
+3\,{w}^{18}+6\,{w}^{17}+{w}^{16}+16\,{w}^{15}+3\,{w}^{14}
+4\,{w}^{13}+5\,{w}^{12}+6\,{w}^{11}+2\,{w}^{10}+9\,{w}^{9}
\nonumber \\
\hspace{-0.95in}&&\quad \quad \quad
+12\,{w}^{8} +4\,{w}^{7}
+11\,{w}^{6}+11\,{w}^{5}+4\,{w}^{4}+4\,{w}^{3}+5\,{w}^{2}+9\,w+10, 
\nonumber 
\end{eqnarray}
\begin{eqnarray}
\label{Rel17r}
\hspace{-0.95in}&&
r_1(w)  \,\, = \, \,\,2\,{w}^{27}+7\,{w}^{26}+15\,{w}^{25}+12\,{w}^{24}
+15\,{w}^{23}+5\,{w}^{22}+7\,{w}^{21}+14\,{w}^{20}+6\,{w}^{19}
\nonumber \\
\hspace{-0.95in}&&\quad \quad \quad
+7\,{w}^{18}
+11\,{w}^{17}+3\,{w}^{16}
+3\,{w}^{15}+4\,{w}^{14}+8\,{w}^{13}+16\,{w}^{12}
\nonumber \\
\hspace{-0.95in}&&\quad \quad \quad
+8\,{w}^{11} +15\,{w}^{10}+15\,{w}^{9}+2\,{w}^{8}+{w}^{7}
\nonumber \\
\hspace{-0.95in}&&\quad \quad \quad
+16\,{w}^{6}+16\,{w}^{5}+7\,{w}^{4}
+6\,{w}^{3}+7\,{w}^{2}+6\,w+7, 
\nonumber \\
\hspace{-0.95in}&&
r_2(w)  \,\, = \, \,\,w \cdot \, 
(12\,{w}^{27}+7\,{w}^{26}+8\,{w}^{25}+11\,{w}^{24}
+13\,{w}^{23}+3\,{w}^{22}+10\,{w}^{21}+14\,{w}^{20}
\nonumber \\
\hspace{-0.95in}&&\quad \quad \quad
+7\,{w}^{19}+6\,{w}^{18}+12\,{w}^{17}
+14\,{w}^{16}+16\,{w}^{15}+16\,{w}^{14}+14\,{w}^{13}+10\,{w}^{12}+4\,{w}^{11}
\nonumber \\
\hspace{-0.95in}&&\quad \quad \quad
+13\,{w}^{10}+7\,{w}^{9}+11\,{w}^{8}+2\,{w}^{7}+8\,{w}^{6}+4\,{w}^{5}+5
\,{w}^{4}+13\,{w}^{3}+7\,{w}^{2}+10\,w+5), 
\nonumber \\
\hspace{-0.95in}&&
r_3(w)  \,\, = \, \,\,{w}^{2} \cdot \, 
(5\,{w}^{26}+9\,{w}^{25}+9\,{w}^{24}+16\,{w}^{23}+4\,{w}^{22}+11\,{w}^{21}
+15\,{w}^{20}+9\,{w}^{19}
\nonumber \\
\hspace{-0.95in}&&\quad \quad \quad
+16\,{w}^{18}+12\,{w}^{17}+3\,{w}^{16}+7\,{w}^{15}
+15\,{w}^{14}+15\,{w}^{13}+11\,{w}^{12}+3\,{w}^{11}+16\,{w}^{10}
\nonumber \\
\hspace{-0.95in}&&\quad \quad \quad
+9\,{w}^{9}+15\,{w}^{8}+9\,{w}^{7}+{w}^{6}+13\,{w}^{5}
+{w}^{4}+5\,{w}^{3}+5\,{w}^{2}+7\,w+1), 
\nonumber 
\end{eqnarray}
\begin{eqnarray}
\hspace{-0.95in}&&
r_4(w)  \,\, = \, \,\,{w}^{3} \cdot \, 
 (15\,{w}^{26}+13\,{w}^{25}+16\,{w}^{24}+13\,{w}^{23}+7\,{w}^{22}+5\,{w}^{21}
+3\,{w}^{20}+2\,{w}^{19}
\nonumber \\
\hspace{-0.95in}&&\quad \quad \quad
+10\,{w}^{18}+10\,{w}^{17}+11\,{w}^{16}+11\,{w}^{15}
+6\,{w}^{13}+16\,{w}^{12}+12\,{w}^{11}+9\,{w}^{10}
\nonumber \\
\hspace{-0.95in}&&\quad \quad \quad
+11\,{w}^{9}+{w}^{8}+2\,{w}^{7}+15\,{w}^{5}+15\,{w}^{4}
+3\,{w}^{3}+15\,{w}^{2}+12\,w+15), 
\nonumber \\
\hspace{-0.95in}&&
r_5(w)  \,\, = \, \,\,{w}^{4} \cdot \, 
(8\,{w}^{25}+4\,{w}^{24}+7\,{w}^{23}+11\,{w}^{22}+4\,{w}^{21}+{w}^{20}
+5\,{w}^{19}
\nonumber \\
\hspace{-0.95in}&&\quad \quad \quad
+15\,{w}^{18}+15\,{w}^{16}+2\,{w}^{15}+{w}^{14}+13\,{w}^{13}
+3\,{w}^{12}+13\,{w}^{11}+11\,{w}^{10}+{w}^{8}
\nonumber \\
\hspace{-0.95in}&&\quad \quad \quad
+5\,{w}^{7}+10\,{w}^{6}
+4\,{w}^{5}+8\,{w}^{4}+16\,{w}^{3}+10\,w+8), 
\nonumber \\
\hspace{-0.95in}&& 
r_6(w)  \,\, = \, \,\,{w}^{5} \cdot \, 
(16\,{w}^{25}+7\,{w}^{24}+2\,{w}^{23}+{w}^{22}+16\,{w}^{21}+12\,{w}^{20}
+16\,{w}^{19}+6\,{w}^{18}
\nonumber \\
\hspace{-0.95in}&&\quad \quad \quad
+10\,{w}^{17}+6\,{w}^{16}+3\,{w}^{15}
+14\,{w}^{14}+16\,{w}^{13}+11\,{w}^{12}+11\,{w}^{11}
+{w}^{10}+15\,{w}^{9}
\nonumber \\
\hspace{-0.95in}&&\quad \quad \quad
+7\,{w}^{8}+16\,{w}^{7}+4\,{w}^{6}+8\,{w}^{5}+14\,{w}^{4}
+7\,{w}^{3}+3\,w+14), 
\nonumber 
\end{eqnarray}
\begin{eqnarray}
\hspace{-0.95in}&&
r_7(w)  \,\, = \, \,\,{w}^{6} \cdot \, (7\,{w}^{24}+4\,{w}^{23}+3\,{w}^{22}
+2\,{w}^{21}+11\,{w}^{20}+15\,{w}^{19}+{w}^{18}
\nonumber \\
\hspace{-0.95in}&&\quad \quad \quad
+3\,{w}^{17}+14\,{w}^{16}
+6\,{w}^{15}+8\,{w}^{14}+6\,{w}^{13}+15\,{w}^{12}+5\,{w}^{11}+3\,{w}^{10}
+8\,{w}^{9}
\nonumber \\
\hspace{-0.95in}&&\quad \quad \quad
+4\,{w}^{8}+15\,{w}^{7}+12\,{w}^{6}+14\,{w}^{4}+14\,{w}^{3}+14), 
\nonumber \\
\hspace{-0.95in}&& 
r_8(w)  \,\, = \, \,\,{w}^{7} \cdot \, 
(2\,{w}^{24}+{w}^{23}+14\,{w}^{22}+4\,{w}^{21}+10\,{w}^{20}
+8\,{w}^{19}+16\,{w}^{18}
\nonumber \\
\hspace{-0.95in}&&\quad \quad \quad
+6\,{w}^{17}+4\,{w}^{16}+10\,{w}^{15}+9\,{w}^{14}
+12\,{w}^{13}+6\,{w}^{12}+14\,{w}^{11}+14\,{w}^{10}
\nonumber \\
\hspace{-0.95in}&&\quad \quad \quad
+16\,{w}^{9}+12\,{w}^{8}
+8\,{w}^{7}+5\,{w}^{6}+4\,{w}^{4}+12\,{w}^{3}+4), 
\nonumber 
\end{eqnarray}
\begin{eqnarray}
\hspace{-0.95in}&& 
r_9(w)  \,\, = \, \,\,{w}^{11} \cdot \, 
(14\,{w}^{20}+{w}^{19}+15\,{w}^{18}+7\,{w}^{17}+10\,{w}^{16}
+6\,{w}^{15} +14\,{w}^{14}
\nonumber \\
\hspace{-0.95in}&&\quad \quad \quad
+10\,{w}^{13}+4\,{w}^{12}+14\,{w}^{11}+11\,{w}^{10}+6\,{w}^{9}
+9\,{w}^{8}+{w}^{7}+9\,{w}^{6}+4\,{w}^{5}
\nonumber \\
\hspace{-0.95in}&&\quad \quad \quad
+10\,{w}^{4}+4\,{w}^{3}+15), 
\nonumber 
\end{eqnarray}
\begin{eqnarray}
\hspace{-0.95in}&&
r_{10}(w)  \,\, = \, \,\,{w}^{12} \cdot \, 
(10\,{w}^{20}+7\,{w}^{19}+5\,{w}^{18}+{w}^{17}+12\,{w}^{16}
+8\,{w}^{15} +16\,{w}^{14}
\nonumber \\
\hspace{-0.95in}&&\quad \quad \quad
+16\,{w}^{13}+15\,{w}^{12}+13\,{w}^{11}+15\,{w}^{10}
+13\,{w}^{9}+4\,{w}^{8}+7\,{w}^{7}+3\,{w}^{6}
\nonumber \\
\hspace{-0.95in}&&\quad \quad \quad
+3\,{w}^{5}+12\,{w}^{4}+11\,{w}^{3}+1), 
\nonumber \\
\hspace{-0.95in}&& 
r_{11}(w)  \,\, = \, \,\,{w}^{16} \cdot \, 
(6\,{w}^{16}+11\,{w}^{15}+5\,{w}^{14}+2\,{w}^{12}+8\,{w}^{11}
+12\,{w}^{10} +16\,{w}^{8}
\nonumber \\
\hspace{-0.95in}&&\quad \quad \quad
+16\,{w}^{7}+14\,{w}^{6}+6\,{w}^{4}+10\,{w}^{3}
+15\,{w}^{2}+7), 
\nonumber 
\end{eqnarray}
\begin{eqnarray}
\label{Rel17rbis}
\hspace{-0.95in}&&
r_{12}(w)  \,\, = \, \,\,{w}^{17} \cdot \, 
(6\,{w}^{16}+16\,{w}^{15}+8\,{w}^{14}+2\,{w}^{12}+7\,{w}^{11}
+9\,{w}^{10} +16\,{w}^{8}
\nonumber \\
\hspace{-0.95in}&&\quad \quad \quad
+14\,{w}^{7}+2\,{w}^{6}+6\,{w}^{4}+13\,{w}^{3}+7\,{w}^{2}+7), 
\nonumber \\
\hspace{-0.95in}&&
r_{13}(w)  \,\, = \, \,\,{w}^{20} \cdot \, 
(6\,{w}^{13}+6\,{w}^{12}+9\,{w}^{9}+11\,{w}^{8}+{w}^{5}+10\,{w}^{4}+7\,w+1), 
\nonumber \\
\hspace{-0.95in}&&
r_{14}(w)  \,\, = \, \,\,{w}^{21} \cdot \, 
(10\,{w}^{13}+13\,{w}^{12}+15\,{w}^{9}+4\,{w}^{8}+13\,{w}^{5}+16\,{w}^{4}+6\,w+5),
\nonumber \\
\hspace{-0.95in}&&
r_{15}(w)  \,\, = \, \,\,{w}^{22} \cdot \, (14\,{w}^{12}+3\,{w}^{8}+12\,{w}^{4}+8),
\nonumber \\
 \hspace{-0.95in}&&
 r_{16}(w)  \,\, = \, \,\,{w}^{23}  \cdot \, (2\,{w}^{12}+15\,{w}^{8}+9\,{w}^{4}+6), 
 \nonumber 
\end{eqnarray}
\begin{eqnarray}
\hspace{-0.95in}&&
r_{17}(w)  \,\, = \, \,\,{w}^{27} \cdot \, ({w}^{2}-w+1) \cdot \, 
 (7\,{w}^{6}+13\,{w}^{5}+10\,{w}^{4}+2\,{w}^{3}+9\,w+9), 
\nonumber \\
\hspace{-0.95in}&&
r_{18}(w)  \,\, = \, \,\,{w}^{28} \cdot \, 
(16\,{w}^{8}+12\,{w}^{7}+2\,{w}^{6}+12\,{w}^{5}+11\,{w}^{4}+5\,{w}^{3}+6), 
\nonumber \\
\hspace{-0.95in}&&
r_{19}(w)  \,\, = \, \,\,{w}^{32} \cdot \, (8\,{w}^{4}+14\,{w}^{3}+7\,{w}^{2}+9),
 \nonumber \\
\hspace{-0.95in}&&
r_{20}(w)  \,\, = \, \,\,{w}^{33}  \cdot \, (15\,{w}^{4}+15\,{w}^{3}+4\,{w}^{2}+2),
\quad \, \, \,   
r_{21}(w)  \,\, = \, \,\,{w}^{36} \cdot \, (5\,w+14),   
 \nonumber \\
 \hspace{-0.95in}&&
r_{22}(w)  \,\, = \, \,\,3\,{w}^{37} \cdot \, (4\,w+3), \quad \quad \,  \, 
r_{23}(w)  \,\, = \, \,\,2\,{w}^{38}, 
\quad \quad \,  \, 
r_{24}(w)  \,\, = \, \,\,w^{39}.
\nonumber 
\end{eqnarray}
One verifies that this polynomial equation is actually satisfied with our series
of $\, 24001$ coefficients modulo $\, p\, = \, 17$.

\vskip .1cm 

\subsection{Polynomial relation for $\, p \, = \, 19$}
\label{pol19}

Modulo $\, p \, = \, \, 19$, we obtained the polynomial relation 
\begin{eqnarray}
\label{Rel19}
\hspace{-0.95in}&& \quad \quad \quad \quad \quad \quad 
\sum_{n=0}^{30} \, \, s_n(w) \cdot S(w)^n \, \, = \, \, \, 0, 
\end{eqnarray}
where:
\begin{eqnarray}
\label{Rel19r}
\hspace{-0.95in}&&
s_{0}(w)  \,\, = \, \,\,7\,{w}^{35}+2\,{w}^{34}+18\,{w}^{33}
+8\,{w}^{32} +{w}^{31}+8\,{w}^{30}+10\,{w}^{29}+9\,{w}^{28}
\nonumber \\
\hspace{-0.95in}&&\quad \quad \quad
+10\,{w}^{27}+16\,{w}^{26}+2\,{w}^{25}+7\,{w}^{24}
+8\,{w}^{23}+2\,{w}^{22}+18\,{w}^{21}+12\,{w}^{19}
\nonumber \\
\hspace{-0.95in}&&\quad \quad \quad
+14\,{w}^{18}+4\,{w}^{17}+12\,{w}^{16}
+13\,{w}^{15}+15\,{w}^{14}+7\,{w}^{13}+8\,{w}^{12}+12\,{w}^{10}
+16\,{w}^{9} 
\nonumber \\
\hspace{-0.95in}&&\quad \quad \quad
+{w}^{7}+15\,{w}^{6}+17\,{w}^{5}+3\,{w}^{4}
+7\,{w}^{3}+6\,{w}^{2}+14\,w +13,
\nonumber 
\end{eqnarray}
\begin{eqnarray}
\hspace{-0.95in}&&
s_{1}(w)  \,\, = \, \,\,10\,{w}^{35}+9\,{w}^{34}+6\,{w}^{33}+{w}^{32}
+9\,{w}^{31}+8\,{w}^{30}+10\,{w}^{28}+17\,{w}^{27}
\nonumber \\
\hspace{-0.95in}&&\quad \quad \quad
+5\,{w}^{26}+7\,{w}^{25}+4\,{w}^{24}+16\,{w}^{22}+15\,{w}^{21}
+9\,{w}^{20}+16\,{w}^{19} +16\,{w}^{18}+11\,{w}^{17}
\nonumber \\
\hspace{-0.95in}&&\quad \quad \quad
+5\,{w}^{16}+5\,{w}^{15}+14\,{w}^{14}+5\,{w}^{13}+13\,{w}^{12}
+3\,{w}^{11}+6\,{w}^{10}+16\,{w}^{9}+17\,{w}^{8}
\nonumber \\
\hspace{-0.95in}&&\quad \quad \quad
+17\,{w}^{7}+11\,{w}^{6} +3\,{w}^{5}
+15\,{w}^{4}+10\,{w}^{3}+4\,{w}^{2}+9\,w+6,
\nonumber \\
\hspace{-0.95in}&&
s_{2}(w)  \,\, = \, \,\,w \cdot \, (3\,{w}^{35}+10\,{w}^{34}+9\,{w}^{33}
+4\,{w}^{32}+5\,{w}^{31}+3\,{w}^{30}+12\,{w}^{29}+5\,{w}^{28}
\nonumber \\
\hspace{-0.95in}&&\quad \quad \quad
+{w}^{27}+5\,{w}^{26}+7\,{w}^{25}+18\,{w}^{23}+9\,{w}^{22}
+2\,{w}^{21}+13\,{w}^{20}
+17\,{w}^{19}+4\,{w}^{18}
\nonumber \\
\hspace{-0.95in}&&\quad \quad \quad
+18\,{w}^{17}+15\,{w}^{16}+{w}^{15}+10\,{w}^{14}+16\,{w}^{13}
+14\,{w}^{12}+17\,{w}^{11}
+18\,{w}^{10}+{w}^{8}
\nonumber \\
\hspace{-0.95in}&&\quad \quad \quad
+2\,{w}^{7}+11\,{w}^{6}
+12\,{w}^{5}+2\,{w}^{4}+13\,{w}^{3}+{w}^{2}+w+3), 
\nonumber 
\end{eqnarray}
\begin{eqnarray}
\hspace{-0.95in}&&
s_{3}(w)  \,\, = \, \,\,{w}^{2} \cdot \, 
(4\,{w}^{34}+11\,{w}^{33}+{w}^{32}+8\,{w}^{31}+5\,{w}^{30}+8\,{w}^{29}
+4\,{w}^{28}+12\,{w}^{27}
\nonumber \\
\hspace{-0.95in}&&\quad \quad \quad
+11\,{w}^{26}+14\,{w}^{25}+9\,{w}^{24}+10\,{w}^{23}+10\,{w}^{22}+3\,{w}^{21}
+11\,{w}^{20}+15\,{w}^{19}
\nonumber \\
\hspace{-0.95in}&&\quad \quad \quad
+7\,{w}^{18}+13\,{w}^{17}+7\,{w}^{16}+18\,{w}^{15}+13\,{w}^{14}+18\,{w}^{13}
+4\,{w}^{12}+4\,{w}^{11}+13\,{w}^{10}
\nonumber \\
\hspace{-0.95in}&&\quad \quad \quad
+7\,{w}^{9}+7\,{w}^{8}+16\,{w}^{7}+{w}^{6}
+3\,{w}^{5}
+7\,{w}^{4}
+18\,{w}^{3}+12\,{w}^{2}+w+8),
\nonumber \\
\hspace{-0.95in}&&
s_{4}(w)  \,\, = \, \,\,{w}^{3} \cdot \, 
(12\,{w}^{34}+4\,{w}^{33}+15\,{w}^{32}+16\,{w}^{31}+18\,{w}^{30}
+16\,{w}^{29}+6\,{w}^{28}
\nonumber \\
\hspace{-0.95in}&&\quad \quad \quad
+18\,{w}^{27}+17\,{w}^{26}+15\,{w}^{25}+12\,{w}^{24}+5\,{w}^{23}+15\,{w}^{22}
+15\,{w}^{21}+5\,{w}^{20}
\nonumber \\
\hspace{-0.95in}&&\quad \quad \quad
+8\,{w}^{19}+18\,{w}^{18}+8\,{w}^{17}+15\,{w}^{15}
+12\,{w}^{14}+17\,{w}^{13}+4\,{w}^{12}
+6\,{w}^{11}+3\,{w}^{10}
\nonumber \\
\hspace{-0.95in}&&\quad \quad \quad
+6\,{w}^{9}+16\,{w}^{8}+10\,{w}^{7}+5\,{w}^{6}+7\,{w}^{5}+14\,{w}^{4}
+6\,{w}^{3}+15\,{w}^{2}+12\,w+11),
\nonumber 
\end{eqnarray}
\begin{eqnarray}
\hspace{-0.95in}&&
s_{5}(w)  \,\, = \, \,\,{w}^{4} \cdot \, (4\,{w}^{33}+11\,{w}^{32}
+10\,{w}^{31} +15\,{w}^{30}+3\,{w}^{29}+2\,{w}^{28}
+4\,{w}^{27}+12\,{w}^{26}
\nonumber \\
\hspace{-0.95in}&&\quad \quad \quad
+3\,{w}^{25}+13\,{w}^{24}+4\,{w}^{23}+11\,{w}^{22}
+7\,{w}^{21}+10\,{w}^{19}+9\,{w}^{18}+9\,{w}^{17}+{w}^{16}
\nonumber \\
\hspace{-0.95in}&&\quad \quad \quad
+2\,{w}^{15}+16\,{w}^{14}+13\,{w}^{13}
+16\,{w}^{12}+18\,{w}^{11}+17\,{w}^{10}+8\,{w}^{9}+18\,{w}^{8}
\nonumber \\
\hspace{-0.95in}&&\quad \quad \quad
+14\,{w}^{5}+4\,{w}^{4}
+5\,{w}^{3}+17\,{w}^{2}+18\,w+18), 
\nonumber \\
\hspace{-0.95in}&&
s_{6}(w)  \,\, = \, \,\,{w}^{5} \cdot \, 
(8\,{w}^{33}+9\,{w}^{32}+5\,{w}^{31}+{w}^{30}+14\,{w}^{29}
+11\,{w}^{28}+5\,{w}^{27}+11\,{w}^{26}
\nonumber \\
\hspace{-0.95in}&&\quad \quad \quad
+10\,{w}^{25}+17\,{w}^{24}+17\,{w}^{23}+17\,{w}^{22}
+2\,{w}^{21}+8\,{w}^{20}+2\,{w}^{18}
+13\,{w}^{17}
\nonumber \\
\hspace{-0.95in}&&\quad \quad \quad
+16\,{w}^{16}+2\,{w}^{15}+2\,{w}^{14}+16\,{w}^{13}+15\,{w}^{12}
+15\,{w}^{11} +7\,{w}^{10}+{w}^{9}+7\,{w}^{8}
\nonumber \\
\hspace{-0.95in}&&\quad \quad \quad
+4\,{w}^{6}+8\,{w}^{5}+2\,{w}^{4}+13\,{w}^{3}+15\,w+11), 
\nonumber 
\end{eqnarray}
\begin{eqnarray}
\hspace{-0.95in}&&
s_{7}(w)  \,\, = \, \,\,{w}^{6}  \cdot \, 
({w}^{32}+18\,{w}^{31}+13\,{w}^{30}+2\,{w}^{29}+13\,{w}^{28}
+13\,{w}^{27}+2\,{w}^{26}+6\,{w}^{25}
\nonumber \\
\hspace{-0.95in}&&\quad \quad \quad
+7\,{w}^{24}+2\,{w}^{23}+13\,{w}^{21}
+12\,{w}^{20}+6\,{w}^{19}+2\,{w}^{18}+11\,{w}^{17}+2\,{w}^{16}
\nonumber \\
\hspace{-0.95in}&&\quad \quad \quad
+2\,{w}^{15}+3\,{w}^{14} +12\,{w}^{13}+5\,{w}^{12}+13\,{w}^{11}
+18\,{w}^{10}+18\,{w}^{9}+3\,{w}^{8}
\nonumber \\
\hspace{-0.95in}&&\quad \quad \quad
+17\,{w}^{6}+3\,{w}^{5}+11\,{w}^{4}+17\,{w}^{3}+4\,w+5), 
\nonumber 
\end{eqnarray}
\begin{eqnarray}
\hspace{-0.95in}&&
s_{8}(w)  \,\, = \, \,\,{w}^{7} \cdot \, 
(11\,{w}^{32}+5\,{w}^{31}+12\,{w}^{30}+2\,{w}^{29}+18\,{w}^{28}
+11\,{w}^{27} +12\,{w}^{26} 
\nonumber \\
\hspace{-0.95in}&&\quad \quad \quad
 +16\,{w}^{25} +8\,{w}^{24}+16\,{w}^{23}+2\,{w}^{22}
+12\,{w}^{20}+9\,{w}^{19}+5\,{w}^{18}
+9\,{w}^{17}
\nonumber \\
\hspace{-0.95in}&&\quad \quad \quad
+18\,{w}^{16}+5\,{w}^{15}+6\,{w}^{14}+16\,{w}^{13}+9\,{w}^{12}
+13\,{w}^{11} +14\,{w}^{10}+14\,{w}^{9}
\nonumber \\
\hspace{-0.95in}&&\quad \quad \quad
+18\,{w}^{8}+14\,{w}^{6}+14\,{w}^{5}+2\,{w}^{4}+4\,{w}^{3}+15), 
\nonumber \\
\hspace{-0.95in}&&
s_{9}(w)  \,\, = \, \,\,{w}^{8}  \cdot \, 
(7\,{w}^{31} +7\,{w}^{30}+16\,{w}^{29}+2\,{w}^{28}+3\,{w}^{27}
+10\,{w}^{26} +18\,{w}^{25} 
\nonumber \\
\hspace{-0.95in}&&\quad \quad \quad 
+5\,{w}^{24}+13\,{w}^{23}+17\,{w}^{22}+12\,{w}^{21}+{w}^{20}+{w}^{19}
+17\,{w}^{18} +17\,{w}^{17}
\nonumber \\
\hspace{-0.95in}&&\quad \quad \quad
+6\,{w}^{16}+15\,{w}^{15}+{w}^{14}+11\,{w}^{13}+10\,{w}^{12}+9\,{w}^{11}
+10\,{w}^{10} +2\,{w}^{9}
\nonumber \\
\hspace{-0.95in}&&\quad \quad \quad
+15\,{w}^{8}+4\,{w}^{6}+18\,{w}^{4}+15\,{w}^{3}+1), 
\nonumber 
\end{eqnarray}
\begin{eqnarray}
\hspace{-0.95in}&&
s_{10}(w)  \,\, = \, \,\,{w}^{12}  \cdot \, 
(16\,{w}^{28}+10\,{w}^{27}+{w}^{26}+13\,{w}^{25}+12\,{w}^{24}+5\,{w}^{23}
\nonumber \\
\hspace{-0.95in}&&\quad \quad \quad
+18\,{w}^{22}
+17\,{w}^{21} +7\,{w}^{20}+4\,{w}^{19}+18\,{w}^{18}+6\,{w}^{17}+16\,{w}^{16}
\nonumber \\
\hspace{-0.95in}&& \quad  \quad  \quad 
+2\,{w}^{15} +18\,{w}^{14}+6\,{w}^{13}+13\,{w}^{12}+2\,{w}^{11}+17\,{w}^{10}+17\,{w}^{9}
\nonumber \\
\hspace{-0.95in}&&\quad \quad \quad
+16\,{w}^{8}+8\,{w}^{7}+9\,{w}^{6}+17\,{w}^{5}
+6\,{w}^{3}+14\,w+16), 
\nonumber \\
\hspace{-0.95in}&&
s_{11}(w)  \,\, = \, \,\,w^{13}  \cdot \, 
(8\,{w}^{27}+13\,{w}^{26}+16\,{w}^{25}+8\,{w}^{24}+2\,{w}^{23}
+14\,{w}^{22}+15\,{w}^{21} 
\nonumber \\
\hspace{-0.95in}&& \quad  \quad  \quad 
 +12\,{w}^{20}   +14\,{w}^{18}+2\,{w}^{17}
+8\,{w}^{16}+14\,{w}^{15}+11\,{w}^{14}+11\,{w}^{13}
+7\,{w}^{12}
\nonumber \\
\hspace{-0.95in}&&\quad \quad \quad
+17\,{w}^{11}+6\,{w}^{9}+18\,{w}^{8}
+4\,{w}^{7}+6\,{w}^{6}+3\,{w}^{5}+12\,{w}^{3}+9),
\nonumber 
\end{eqnarray}
\begin{eqnarray}
\label{Rel19rter2}
\hspace{-0.95in}&&
s_{12}(w)  \,\, = \, \,\,{w}^{14}  \cdot \, 
(8\,{w}^{27}+{w}^{26}+4\,{w}^{25}+18\,{w}^{24}+9\,{w}^{22}
+3\,{w}^{21}+15\,{w}^{20}
\nonumber \\
\hspace{-0.95in}&&\quad \quad \quad
+6\,{w}^{18}+6\,{w}^{17}+2\,{w}^{16}+3\,{w}^{15}+3\,{w}^{13}
+9\,{w}^{12}+7\,{w}^{11}
+6\,{w}^{9}+16\,{w}^{8}
\nonumber \\
\hspace{-0.95in}&&\quad \quad \quad
+{w}^{7}+4\,{w}^{6}+10\,{w}^{3}+9), 
\nonumber \\
\hspace{-0.95in}&&
s_{13}(w)  \,\, = \, \,\,{w}^{18} \cdot \, 
(15\,{w}^{23}+6\,{w}^{22}+12\,{w}^{21}+{w}^{19}+{w}^{18}+2\,{w}^{17}
+14\,{w}^{14}+3\,{w}^{13}
\nonumber \\
\hspace{-0.95in}&&\quad \quad \quad
+2\,{w}^{12}+13\,{w}^{10}+3\,{w}^{9}+6\,{w}^{8}+12\,{w}^{5}
+11\,{w}^{4}+9\,{w}^{3}+16), 
\nonumber \\
\hspace{-0.95in}&&
s_{14}(w)  \,\, = \, \,\,{w}^{19}  \cdot \, 
(2\,{w}^{23}+10\,{w}^{22}+8\,{w}^{21}+18\,{w}^{18}+15\,{w}^{17}
+12\,{w}^{14}+5\,{w}^{13}
\nonumber \\
\hspace{-0.95in}&&\quad \quad \quad
+14\,{w}^{12}+16\,{w}^{9}+7\,{w}^{8}+13\,{w}^{5}+12\,{w}^{4}+6\,{w}^{3}+3),
\nonumber 
\end{eqnarray}
\begin{eqnarray}
\hspace{-0.95in}&&
s_{15}(w)  \,\, = \, \,\,{w}^{20} \cdot \, 
(16\,{w}^{22}+18\,{w}^{21}+3\,{w}^{18}+17\,{w}^{17}
+8\,{w}^{13}+3\,{w}^{12}+9\,{w}^{9}
\nonumber \\
\hspace{-0.95in}&&\quad \quad \quad
+13\,{w}^{8}+4\,{w}^{4}+4\,{w}^{3}+10), 
\nonumber 
\end{eqnarray}
\begin{eqnarray}
\hspace{-0.95in}&&
s_{16}(w)  \,\, = \, \,\,{w}^{24} \cdot \, 
(12\,{w}^{19}+12\,{w}^{18}+16\,{w}^{14}+6\,{w}^{10}
+2\,{w}^{9}+10\,{w}^{5}+3\,w+9), 
\nonumber \\
\hspace{-0.95in}&&
s_{17}(w)  \,\, = \, \,\,2\,{w}^{25} \cdot \, 
(4\,{w}^{18}+9\,{w}^{14}+7\,{w}^{9}+8\,{w}^{5}+3), 
\nonumber \\
\hspace{-0.95in}&&
s_{18}(w)  \,\, = \, \,\,{w}^{26} \cdot \, (18\,{w}^{18}+3\,{w}^{9}+4), 
\nonumber \\
\hspace{-0.95in}&&
s_{19}(w)  \,\, = \, \,\,{w}^{30} \cdot \, 
(13\,{w}^{14}+15\,{w}^{13}+{w}^{12}+17\,{w}^{11}+18\,{w}^{10}
+17\,{w}^{8}+4\,{w}^{7}
\nonumber \\
\hspace{-0.95in}&&\quad \quad \quad
+15\,{w}^{6}+17\,{w}^{5}+14\,{w}^{4}+8\,{w}^{3}+14\,w+8), 
\nonumber
\end{eqnarray}
\begin{eqnarray}
\label{Rel19rter}
\hspace{-0.95in}&&
s_{20}(w)  \,\, = \, \,\,{w}^{31} \cdot \, 
 (4\,{w}^{14}+8\,{w}^{13}+3\,{w}^{12}+3\,{w}^{11}+15\,{w}^{10}
\nonumber \\
\hspace{-0.95in}&&\quad \quad \quad
+4\,{w}^{9}+18\,{w}^{8} +3\,{w}^{7}
\nonumber \\
\hspace{-0.95in}&&\quad \quad \quad
+15\,{w}^{6}+14\,{w}^{5}+10\,{w}^{4}+5\,{w}^{3}+2), 
\nonumber \\
\hspace{-0.95in}&& 
s_{21}(w)  \,\, = \, \,\,{w}^{32} \cdot \,  
({w}^{13}+15\,{w}^{12}+13\,{w}^{11}+17\,{w}^{10}+3\,{w}^{9}+{w}^{8}
+14\,{w}^{7}+3\,{w}^{6}
\nonumber \\
\hspace{-0.95in}&& \quad \quad \quad
+6\,{w}^{4}+6\,{w}^{3}+12),
\nonumber \\
\hspace{-0.95in}&&
s_{22}(w)  \,\, = \, \,\,{w}^{36} \cdot \, 
 (4\,{w}^{10}+9\,{w}^{9}+13\,{w}^{8}+{w}^{7}
+15\,{w}^{6}+8\,{w}^{5}
+15\,{w}^{4}
\nonumber \\
\hspace{-0.95in}&& \quad \quad \quad
+15\,{w}^{3}+5\,w+15), 
\nonumber \\
\hspace{-0.95in}&&
s_{23}(w)  \,\, = \, \,\,{w}^{37} \cdot \, 
 (16\,{w}^{9}+4\,{w}^{8}+18\,{w}^{7}+18\,{w}^{6}
+5\,{w}^{5}+2\,{w}^{4}+16\,{w}^{3}+14), 
\nonumber 
\end{eqnarray}
\begin{eqnarray}
\hspace{-0.95in}&& 
s_{24}(w)  \,\, = \, \,\, {w}^{38}  \cdot \, 
(8\,{w}^{9}+6\,{w}^{8}+7\,{w}^{7}+6\,{w}^{6}
+2\,{w}^{4}+13\,{w}^{3}+7), 
\nonumber \\
\hspace{-0.95in}&&
s_{25}(w)  \,\, = \, \,\,{w}^{42} \cdot \, 
(5\,{w}^{5}+18\,{w}^{4}+15\,{w}^{3}+4\,w+2), 
\nonumber \\
\hspace{-0.95in}&&
s_{26}(w)  \,\, = \, \,\,{w}^{43} \cdot \, 
 (14\,{w}^{5}+3\,{w}^{4}+{w}^{3}+15), 
\nonumber \\
\hspace{-0.95in}&&
s_{27}(w)  \,\, = \, \,\,{w}^{44}  \cdot \, 
(9\,{w}^{4}+6\,{w}^{3}+13), 
\nonumber \\
\hspace{-0.95in}&&
s_{28}(w)  \,\, = \, \,\,{w}^{48} \cdot \, (12\,w+5),
 \quad \quad \quad  
s_{29}(w)  \,\, = \, \,\,12\,{w}^{49}, \quad \quad  \quad 
s_{30}(w)  \,\, = \, \,\,{w}^{50}. 
\nonumber 
\end{eqnarray}
One verifies that this polynomial equation is actually satisfied 
with $\, 23756$ coefficients
of our series modulo $\, p\, = \, 19$.

\vskip .1cm 

\vskip .1cm 

\section{ Reduction of hypergeometric functions}
\label{appendixhyper}

\subsection{ Reduction of $\, _nF_{n-1}$ hypergeometric functions modulo primes}
\label{appendixhyperapp}

Let us consider the series expansions (with integer coefficients)
 of the hypergeometric function  
$\, _4F_3([1/2, \,1/2, \,1/2, \,1/2], \, [1, \, 1, \, 1], 256 \, x)$,
 which corresponds
to a {\em Calabi-Yau operator}~\cite{CalabiYauIsing,bridged}.  It 
is the diagonal of a rational function~\cite{Short,Big} since 
it is the {\em Hadamard product}~\cite{Short} of four times 
the algebraic function $\, (1\, -\,4 \, x)^{-1/2}$. 
This ensures that this series reduces to an algebraic function 
modulo any prime~\cite{Short,Big} (or power of prime).

Let us perform the same calculations as in sections \ref{reduc}
and \ref{algmodprime}.
The series reads: 
 \begin{eqnarray}
\label{4F3}
\hspace{-0.95in}&& \, 
 _4F_3\Bigl([{{1} \over {2}}, \,{{1} \over {2}}, \,{{1} \over {2}}, \,{{1} \over {2}}], 
\, [1, \, 1, \, 1], 256 \, x\Bigr) \, = \, \, \,   
1 \,\,   +16\,x\,  +1296\,{x}^{2} +160000\,{x}^{3}+24010000\,{x}^{4}\,
\nonumber \\
\hspace{-0.95in}&& \quad \quad \quad +4032758016\,{x}^{5}
\,\, +728933458176\,{x}^{6}\,\, +138735983333376\,{x}^{7}
\,\,\,  + \,\,\, \cdots 
\end{eqnarray}

The reduction of this hypergeometric series is a very simple algebraic 
function of the form $\, P(x)^{-1/N}$ where $\, N$ is an integer
and where $\, P(x)$ is a polynomial, which corresponds to the truncation of the 
series expansion of the hypergeometric series modulo the prime $\, p$.

For instance, modulo $\, 23$, the hypergeometric function (\ref{4F3})
becomes the algebraic function $\,\, 1/P(x)^{1/22}$, where 
 the polynomial $\, P(x)$ reads:
\begin{eqnarray}
\label{4F3p23}
\hspace{-0.95in}&&   \,   
P(x) \, \, = \, \, \, 
\Bigl(\, 
 _4F_3\Bigl([{{1} \over {2}}, \,{{1} \over {2}}, \,{{1} \over {2}}, \,{{1} \over {2}}], 
\, [1, \, 1, \, 1], 256 \, x\Bigr)\Bigr)^{-22} 
\qquad \quad    \bmod \, \, \,  23
  \\
\hspace{-0.95in}&&    \quad   \, \, = \, \, \, 
1\,  \, +16\,x \,+8\,{x}^{2}+12\,{x}^{3}+{x}^{4}+{x}^{5}+3\,{x}^{6}+4\,{x}^{7}\, 
+18\,{x}^{8}+16\,{x}^{9}+12\,{x}^{10}+{x}^{11}
  \nonumber
\end{eqnarray}
More generally one can conjecture that, modulo almost all prime $\, p$,
 the hypergeometric series
to the power $\, -(p-1)$ is a polynomial:
\begin{eqnarray}
\label{4F3pp}
\hspace{-0.95in}&&   \quad  \, \, \,  
P(x) \, \, = \, \, \, 
\Bigl(\,  
 _4F_3\Bigl([{{1} \over {2}}, \,{{1} \over {2}}, \,{{1} \over {2}}, \,{{1} \over {2}}], 
\, [1, \, 1, \, 1], 256 \, x\Bigr)\Bigr)^{-(p\, -1)} 
\quad  \quad   \bmod \, \, \,  p.
\end{eqnarray}
This polynomial is of degree  $\,98 $ for the prime $\, 197$, 
of  degree $\, 411$ for the prime $\, 823$, of degree $\, 1121$ for 
 the prime $\,2243$.
One can conjecture, modulo almost all prime $\, p$, that 
the degree of this polynomial is  $\, (p\, -1)/2$.

\vskip .2cm 

{\bf Remark:}  One remarks that the polynomial $\, P(x)$ 
corresponds to a truncation of the hypergeometric function we started from. For 
instance, modulo $\, p\, = \, 23$, the series expansion of the 
$\, _4F_3$ hypergeometric function reads:
\begin{eqnarray}
\label{4F3p23dd}
\hspace{-0.95in}&&    
  _4F_3\Bigl([{{1} \over {2}}, \,{{1} \over {2}}, \,{{1} \over {2}}, \,{{1} \over {2}}], 
\, [1, \, 1, \, 1], 256 \, x\Bigr)
\nonumber \\
\hspace{-0.95in}&&    \quad  \, \, = \, \, \,  \, 
1\, \, \, +16\,x\, \, +8\,{x}^{2}+12\,{x}^{3}
+{x}^{4}+{x}^{5}+3\,{x}^{6}+4\,{x}^{7}\, 
+18\,{x}^{8} +16\,{x}^{9} +12\,{x}^{10}+{x}^{11} 
\nonumber \\
\hspace{-0.95in}&&    \quad  \, \,    \quad \quad \quad 
 +16\,{x}^{23} +3\,{x}^{24}+13\,{x}^{25}+8\,{x}^{26} \, +16\,{x}^{27}
 \,\,\,  \,  + \, \cdots \quad \quad    \bmod \, \, \, 23.
\nonumber 
\end{eqnarray}
\begin{eqnarray}
\hspace{-0.95in}&&    \quad  \, \, = \, \, \,  \, 
\Bigl(\, 
 _4F_3\Bigl([{{1} \over {2}}, \,{{1} \over {2}}, \,{{1} \over {2}}, \,{{1} \over {2}}], 
\, [1, \, 1, \, 1], 256 \, x\Bigr)\Bigr)^{-22} \,  \, \, \nonumber \\
\hspace{-0.95in}&&    \quad  \, \,    \quad \quad \quad 
 +16\,{x}^{23} +3\,{x}^{24}+13\,{x}^{25}+8\,{x}^{26} \, +16\,{x}^{27}
 \,\,\,  \,  + \, \cdots \quad \quad    \bmod \, \, \, 23.
\end{eqnarray}
which corresponds to the fact that: 
\begin{eqnarray}
\label{4F3p23dd}
\hspace{-0.95in}&& \quad    \quad    \quad \quad 
  _4F_3\Bigl([{{1} \over {2}}, \,{{1} \over {2}}, \,{{1} \over {2}}, \,{{1} \over {2}}], 
\, [1, \, 1, \, 1], 256 \, x\Bigr)^{23} \, - \, 1 
 \\
\hspace{-0.95in}&&    \quad \quad  \, \,  
 \quad    \quad   \quad   \quad    \, \, = \, \, \, \, \, 
16\,{x}^{23} \, +8\,{x}^{46}+12\,{x}^{69}+{x}^{92} \, \, 
 \, + \, \cdots \quad\quad     \bmod \, \, \, 23.
\nonumber
\end{eqnarray}
More generally, one has:
\begin{eqnarray}
\label{4F3p23dd}
\hspace{-0.95in}&&  \,   \,
 _4F_3\Bigl([{{1} \over {2}}, \,{{1} \over {2}}, \,{{1} \over {2}}, \,{{1} \over {2}}], 
\, [1, \, 1, \, 1], 256 \, x\Bigr)^{M} \, - \, 1 
\, \, \, = \, \, \, \, \, 16\,M \cdot \, x \, \,
+16\,M \cdot \, \left( 8\,M+73 \right) \cdot \, {x}^{2} \, 
 \nonumber  \\
\hspace{-0.95in}&&     \quad  \quad  \quad   \quad  \quad  
+{{256 } \over {3}} \cdot \,  M \cdot \, (8\,{M}^{2}+219\,M+1648) \cdot \,{x}^{3} \, 
 \,\, \,\,  + \, \, \cdots
\end{eqnarray}
all the coefficients of this series (\ref{4F3p23dd}) are of 
the form $\, M \cdot \, P(M)/d$ where $\, P(M)$ is a polynomial 
with integer coefficients, the denominator $\, d$ is an 
integer. Modulo $\, M$  the coefficients of this expansion are all equal 
to zero, except when the denominator of this coefficient is divisible by $\, M$.

\vskip .1cm 

\subsection{Reduction of hypergeometric functions modulo power of primes}
\label{appendixhypermorepower}

The algebraic expressions, corresponding to 
reductions of hypergeometric functions modulo {\em power of primes}, 
are much more complicated.  Let us just consider 
the previous $\, _4F_3$ hypergeometric function, for instance,
modulo $\, 3^2$. This series modulo $\, 3^2$ reads: 
\begin{eqnarray}
\label{4F3p23ddgg}
\hspace{-0.95in}&& \quad \quad \,
S \,\, = \, \,\,\,1\,\,+7\,x\,\, +7\,{x}^{3}\,+7\,{x}^{4}\,+7\,{x}^{9}\,
+4\,{x}^{10}\,+7\,{x}^{12}\,+7\,{x}^{13}\,+7\,{x}^{27}\,+4\,{x}^{28}\,
\nonumber \\
\hspace{-0.95in}&& \quad \quad \quad \quad \quad \quad 
+4\,{x}^{30} \,+4\,{x}^{31}\,+7\,{x}^{36}\,+4\,{x}^{37}\,+7\,{x}^{39}\,
+7\,{x}^{40}\,\,\, + \,\, \, \cdots 
\end{eqnarray}
It is solution of the polynomial relation 
\begin{eqnarray}
\label{4F3p23ddgg}
\hspace{-0.95in}&& \quad \quad \quad \quad 
(x^7\,+2\,x^6\,+x^5\,+x^2\,+2\,x\,+1) \cdot \, S^4 
 \\
\hspace{-0.95in}&& \quad \quad \quad \quad  \quad \quad \,\, \,  \, 
+(x^6\,+x^5\,+x\,+1) \cdot \, S^2 \, \, \, 
+7 \cdot \, (1+x^5) \,\, \, = \, \, \,\, 0  
\quad \quad  \, \,  \bmod \,\, 3^2.
\nonumber 
\end{eqnarray}

\vskip .1cm 

\subsection{ More reduction of hypergeometric functions}
\label{appendixhypermore}

Such result generalizes to other hypergeometric functions. For instance 
for the $\, _5F_4$  hypergeometric functions: 
\begin{eqnarray}
\label{5F4p5}
\hspace{-0.95in}&&    \quad  \quad 
P(x) \, \, = \, \, \, \, 1 \, \, + 2\, x \, +\, x^2 
 \\
\hspace{-0.95in}&&  \quad  \quad  \quad \quad 
 \, \, = \, \, \, 
\Bigl(\,
 _5F_4\Bigl([{{1} \over {2}}, \,{{1} \over {2}},
 \,{{1} \over {2}}, \,{{1} \over {2}}, \,{{1} \over {2}}], 
\, [1, \, 1, \, 1, \, 1], \,2^{10} \, x\Bigr) \Bigr)^{-4} 
\qquad    \,  \bmod \, \,  5,
 \nonumber 
\end{eqnarray}
\begin{eqnarray}
\label{5F4p5other}
\hspace{-0.95in}&&    \quad  \quad 
P(x) \, \, = \, \, \, \, 1 \, \, + 2\, x \, +\, 5\, x^2 
 \\
\hspace{-0.95in}&&  \quad  \quad  \quad \quad 
 \, \, = \, \, \, 
\Bigl(\, _5F_4\Bigl([{{1} \over {2}}, \,{{1} \over {2}}, \,{{1} \over {2}}, 
\,{{1} \over {3}}, \,{{1} \over {3}}], 
\, [1, \, 1, \, 1, \, 1], \, 2^{6} \, 3^4  \, x\Bigr)
\Bigr)^{-6} 
\qquad    \bmod \, \,  7, \nonumber 
\end{eqnarray}
but 
\begin{eqnarray}
\label{5F4p5other}
\hspace{-0.95in}&&    \quad  \quad 
P(x) \, \, = \, \, \,  
1\,  \, +2\,x\, \, +4\,{x}^{2}\, +3\,{x}^{5}\, +{x}^{6}\, +2\,{x}^{7}
 \\
\hspace{-0.95in}&&  \quad  \quad  \quad \quad 
 \, \, = \, \, \, 
 \, \Bigl(\,  _5F_4\Bigl([{{1} \over {2}}, \,{{1} \over {2}}, \,{{1} \over {2}}, 
\,{{1} \over {3}}, \,{{1} \over {3}}], 
\, [1, \, 1, \, 1, \, 1], \, 2^{6} \, 3^4  \, x\Bigr)
\Bigr)^{-24} 
\qquad   \,   \bmod \, \,  5. \nonumber 
\end{eqnarray}
In fact the hypergeometric series, modulo $\, p$, are
of the form $\, P(x)^{-1/N}$ where $\, N$ is an integer,
 not necessarily
equal to $\, -(p-1)$, which is such that 
$\, -N \, = 1 \quad \bmod \, \, p$.

\vskip .2cm

\subsection{Reductions modulo primes of 
                  $\, _3F_2([1/9, \, 4/9, \, 5/9],[1/3, \, 1], 3^6 \, x)$ \\}
\label{holoreduc}

Let us now consider the $\, _3F_2$ hypergeometric function 
$\, _3F_2([1/9, \, 4/9, \, 5/9],[1/3, \, 1], 3^6 \, x)$. This 
hypergeometric function has a series expansion 
{\em with integer coefficients}:
\begin{eqnarray}
\label{3F2}
\hspace{-0.95in}&& 
  _3F_2\Bigl([{{1} \over {9}}, \,{{4} \over {9}}, \,{{5} \over {9}}], 
\, [{{1} \over {3}}, \, 1], 3^6 \, x\Bigr)
\, \, = \, \, \,  \,
1 \,  \, \, +60\,x  \, \, +20475\,{x}^{2} \, +9373650\,{x}^{3} \, 
\nonumber \\
\hspace{-0.95in}&& \quad 
 +4881796920\,{x}^{4}  
 +2734407111744\,{x}^{5} 
  +1605040007778900\,{x}^{6}
  \nonumber \\
\hspace{-0.95in}&& \quad 
+973419698810097000\,{x}^{7} \,  +   \, \cdots 
\end{eqnarray} 

This  $\, _3F_2$ hypergeometric function has been introduced by 
G. Christol~\cite{Short,Big,Christol}, a few decades ago, 
to provide an example of holonomic $\, G$-series 
with {\em integer coefficients} that may not be a diagonal 
of a rational function (it is still an open question to see whether 
this function is, or is not, the diagonal of rational function). 

If this hypergeometric function were the diagonal of a rational function
it would reduce to  algebraic functions modulo every prime, in particular
small primes like $\, 2$, $\, 3$, $\, 5$, $\, 7$. Considering the
series (\ref{3F2}) modulo these primes, in order to see whether they 
reduce, or not, to algebraic functions modulo these primes, is certainely
worth doing to have a better hint on the very nature of
this  hypergeometric function: diagonal of rational function, or not. 
 
Considering the previous series expansion with integer 
coefficients (\ref{3F2}), modulo the prime $\, 2$, we obtained 
a (quite lacunary) series of the first $\, 533000$ coefficients: 
\begin{eqnarray}
\label{3F2mod2}
\hspace{-0.95in}&&  \, 
S \, = \, \, \, \, \,  1 \, \,  \,+{x}^{2} \,\, +{x}^{128} \, \,+{x}^{130} \, \,
+{x}^{8192}+{x}^{8194} \, \,+{x}^{8320} \,\, +{x}^{8322}
 \\
\hspace{-0.95in}&& \quad \, 
+{x}^{524288} +{x}^{524290} +{x}^{524416} +{x}^{524418} +{x}^{532480}
+{x}^{532482} +{x}^{532608} \, + \, \, O(x^{533000}) 
\nonumber 
\end{eqnarray}
In contrast with the calculations  performed in sections \ref{reduc}
and \ref{algmodprime}, or in the previous section (\ref{appendixhyperapp}),
it becomes hard to find the polynomial relation (if it exists !) 
this series (\ref{3F2mod2})
satisfies, even modulo $\, 2$. The reason is that the series (\ref{3F2mod2}) 
satisfies\footnote[1]{We thank A. Bostan for kindly providing this result.},
 modulo $\, 2$, an algebraic relation of slightly large degree 
$\,2^6\, -1 \, = \, \,  63$, namely
 $\, \, (1+x^2)\cdot \, S^{63} \, - 1 \,=  \, 0$.
One can check directly that: 
\begin{eqnarray}
\label{63lindirect}
\hspace{-0.95in}&& \quad  \quad  \quad  \quad
 _3F_2\Bigl([{{1} \over {9}}, \,{{4} \over {9}}, \,{{5} \over {9}}], 
\, [{{1} \over {3}}, \, 1], 3^6 \, x\Bigr)
\, \, \,  = \,  \, \, \, 
(1 \, + \, x^2)^{-1/63} \qquad \bmod \, \, 2. 
\end{eqnarray}

\vskip .1cm 

The series (\ref{3F2}) becomes trivial modulo the prime $\, 3$,
 however, if one considers, instead, the series 
$\, \,S \, = \,  $
$1\, +(_3F_2([1/9, \, 4/9, \, 5/9],[1/3, \, 1], 3^6 \, x) \, -1)/15$,
this series expansion, modulo $\, 3$, is the 
lacunary series $ \,1 \, + \,  \sum \, x^{3^n}$: 
\begin{eqnarray}
\label{3F2mod3}
\hspace{-0.95in}&& \,  
1 \, +x \, +x^3\, +x^9\, +x^{27}\, +x^{81}\, +x^{243} 
\, +x^{729} \, +x^{2187} \, +x^{6561} \, +x^{19683}
 \,   + \, \, \cdots 
\end{eqnarray}
which is algebraic since it 
satisfies, modulo $\, 3$, the polynomial relation 
$\, \,\,  S^3 \,  +x \, = \, S$. 

\vskip .3cm 

{\bf Remark:} Even in a holonomic framework, the property to reduce to an 
algebraic function modulo every prime (and power of prime) is probably 
more general than being the diagonal of a rational function. 
For holonomic $\, G$-series with integer coefficients that do not reduce 
to diagonal of a rational function, one must not seek for polynomial relations 
$\, P(x, \, S) \, \, = \, \, \, 0$ where the degrees in $\, x$ and $\, S$ 
are not too drastically different, but one must rather seek for  polynomial 
relations of the ``Frobenius'' type:
\begin{eqnarray}
\label{Frobenius}
\hspace{-0.95in}&& \quad \quad \quad \quad \quad \quad \quad 
\sum a_i(x) \cdot \, S^{p^i}  
\,  \,=\, \,  \, 0   \quad \quad  \quad \quad   \bmod \, \, p
\end{eqnarray}
where the degree in $\, S$, namely $\, p^N$ for some $\, N$ 
integer, can be quite large.

\vskip .1cm 

Modulo $\, 5$ the series  (\ref{3F2}) becomes a function of the 
variable\footnote[9]{Sometimes called ``constant'' by some authors because 
it derivative is $\, 5 \cdot x^4$ which is zero mod. $\, 5$.} $\, x^5$: 
\begin{eqnarray}
\label{3F2mod5}
\hspace{-0.95in}&& \, \,
1\, \, \, +4\,{x}^{5}\,\, +2\,{x}^{10}\,+3\,{x}^{25}\,+2\,{x}^{30}\,+2\,{x}^{35}
+2\,{x}^{50}\,+3\,{x}^{55}\,+4\,{x}^{250}\, +{x}^{255}
 \\
\hspace{-0.95in}&& \quad    \quad \,
+3\,{x}^{260} \,+2\,{x}^{275}\,+3\,{x}^{280}\,+3\,{x}^{285}
+3\,{x}^{300}  \,+ 2\,{x}^{305}\,+{x}^{375}\,+4\,{x}^{380}\, 
\,\, \, + \, \, \cdots 
\nonumber
\end{eqnarray}
For this series (\ref{3F2mod5}), as well as the reduction of (\ref{3F2})
modulo $\, 7$, it is extremely hard to see whether these series 
satisfy a polynomial relation, 
{\em even of the Frobenius type} (\ref{Frobenius}). 

\vskip .1cm 

\section{Ratio of holonomic functions versus  ratio 
of diagonal rational functions\\}
\label{versus}

 Let us consider a quite pedagogical and important example related to 
the {\em theory of elliptic curves}, and the concept of 
{\em mirror maps}~\cite{CalabiYauIsing,Singularities}.

Let us consider $\, \tau \, = \, -\, \pi\, \rho $ the ratio 
 of the two periods of an elliptic function as a function 
of the lambda modulus $\, \lambda \, = \, \, k^2$:
\begin{eqnarray}
\label{tau}
\hspace{-0.75in}&& \quad \quad \quad \quad \quad \quad
\rho \, \,\,  = \, \, \,  \,\,
{{  _2F_1\Bigl([{{1} \over {2}},\,{{1} \over {2}}],[1],\, 1 \, -\, k^2 \Bigr)
} \over { 
 _2F_1\Bigl([{{1} \over {2}},\,{{1} \over {2}}],[1],\, k^2 \Bigr) 
}}, 
\end{eqnarray}
where the complete elliptic integral of the first kind
and the complementary complete elliptic integral  of the first kind
have the series expansions 
\begin{eqnarray}
\label{first}
\hspace{-0.95in}&& \quad \, \,\,
 _2F_1\Bigl([{{1} \over {2}},\,{{1} \over {2}}],[1],\, x^2 \Bigr)
\, = \, \,\,\, \, 1 \,\, \,  +{\frac {x^2}{4}} \,\, 
 +{\frac {9}{64}}{x}^{4} \, \, +{\frac {25}{256}}{x}^{6}
 \, \, +{\frac {1225}{16384}}{x}^{8}\,\,\, \,    + \,\, \cdots  
\end{eqnarray}
and 
\begin{eqnarray}
\label{complementary}
\hspace{-0.95in}&& 
 _2F_1\Bigl([{{1} \over {2}},\,{{1} \over {2}}],[1],\, 1 \, -x^2 \Bigr)
\,\,  = \, \, \, \, \,  
\ln(x) \cdot \, \,
  _2F_1\Bigl([{{1} \over {2}},\,{{1} \over {2}}],[1],\, k^2 \Bigr)
\, \,\,  + \, \, y_0(x) \qquad \quad  \quad \hbox{where:} 
\nonumber \\
\hspace{-0.95in}&& \quad 
y_0(x) \, = \, \, \, \,  \, 
 {{x^2} \over {4}} \, \,\, \,    +{\frac {21\,{x}^{4}}{128}}
\, \,    \, +{\frac {185\,{x}^{6}}{1536}}
\,\,  +{\frac {18655\,{x}^{8}}{196608}}
 \,\,  +{\frac {102501\,{x}^{10}}{1310720}} \, \, 
+{\frac {1394239\,{x}^{12}}{20971520}}
\nonumber \\
\hspace{-0.95in}&& \quad \quad \quad \quad\quad 
 \, +{\frac {33944053\,{x}^{14}}{587202560}} \, 
+{\frac {3074289075\,{x}^{16}}{60129542144}} \,
 +{\frac {99205524275\,{x}^{18}}{2164663517184}}
\,\,\,   + \,\, \cdots 
\end{eqnarray}
Introducing the two second order linear differential 
operators (here $D_x \, = \, \, d/dx$)
\begin{eqnarray}
\label{firstoper}
\hspace{-0.95in}&& \quad \quad \quad \quad
L_2 \, \, = \, \, \, \,
 (x^2-1)\cdot \, x \cdot \, D_x^2
 \,  \, \, + \, \, (3\, x^2-1)\cdot \, D_x\, \, + \, \,x, 
\\
\hspace{-0.95in}&& \quad \quad \quad \quad
M_2 \, \, = \, \, \, \,
 (x^2-1)\cdot \, x^2 \cdot \, D_x^2 
\,  \, \, + \, \, (3\, x^2-1)\cdot \, x \cdot \, D_x\,  \,+ \, \,1, 
\end{eqnarray} 
the complete elliptic integral of the first kind (\ref{first})
 is solution of $\, L_2$ when 
 the series $\, y_0(x)$ in (\ref{complementary}) is solution
of the fourth-order linear differential operator 
$\, L_4 \, = \, \, M_2 \cdot L_2$. 
Therefore the ratio $\, \rho$ in (\ref{tau}) reads
\begin{eqnarray}
\label{taubis}
\hspace{-0.95in}&& \quad 
\rho \, \,\,  = \, \, \,  \ln(x) \, \, + \, \, \, r(x)   \qquad \, \, 
\hbox{where:}  \qquad \, \, 
 r(x) \,\,  = \, \,  
{{  y_0} \over { 
 _2F_1\Bigl([{{1} \over {2}},\,{{1} \over {2}}],[1],\, x^2 \Bigr)
}}.
\end{eqnarray}
It is well-known that the ratio $\, \tau$ (and thus the ratio $\, \rho$)
satisfies a very simple {\em non-linear "Schwarzian differential equation"}:
\begin{eqnarray}
\label{Schwarzfirst}
\hspace{-0.75in}&& \quad \quad \quad \quad \quad \quad
\{ \rho, \, \, \lambda \}   \,  \, \, = \, \, 
\,  \, \, \,
{{1} \over {2}} \cdot \, 
 {{ (\lambda^2  \, - \,\lambda \,+1) } \over { \lambda^2 \cdot (\lambda\, -1)^2 }}, 
\end{eqnarray}
where, if $\, x$ is the modulus $\, k$ of elliptic function, where
 $\, \lambda$ denotes the ``lambda modulus''
 $\, \lambda \, = \, \, k^2 \, = \, \,x^2$, 
and where $\, \{ \rho, \, \, \lambda \}$ denotes the {\em Schwarzian derivative}. 
 
From (\ref{taubis}) and (\ref{Schwarzfirst}) one immediately finds 
that $\, r(x)$, the ratio
of two holonomic functions, satisfy a non-linear differential 
equation, that we will not write here.

In order to have series with integer coefficients, let us scale $\, x$ 
by a factor $\, 4$: $\, x \, \rightarrow \, \, 4 \, x$. The elliptic 
integral (\ref{first}), which is a 
diagonal of a rational function,
has very simple reductions modulo primes. 
For instance, modulo $\, p \, = \, 7$, it reads:
\begin{eqnarray}
\label{reduc7tt}
\hspace{-0.95in}&& \quad  \,  \, \, \, \,  
_2F_1\Bigl([{{1} \over {2}}, \,{{1} \over {2}}], \, [1], \, \, 16 \, x^2 \Bigr) 
\, \, = \, \, \,  (1 \, +4\,{x}^{2}+\,{x}^{4}+{x}^{6})^{-1/6}
\qquad \, \,  \bmod \, \,  7.
\end{eqnarray}

Unfortunately one {\em cannot define the reduction of the holonomic series} 
$\, y_0$, solution of a fourth-order 
linear differential operator. One sees that this series (even with a 
rescaling $\, x\, \, \rightarrow \, \, 4 \, x$, or
even any rescaling by an integer, cannot be recast into 
a series with integer coefficients: 
{\em it is not globally bounded}~\cite{Short,Big}.
In the denominators of the successive coefficients of this series 
almost every prime occurs, thus, one cannot look at this series 
modulo a prime\footnote[2]{In Maple the mod prime command gives a 
``Error, the modular inverse does not exist'' warning.}.

\section{Non-linear differential equation for a  ratio 
of diagonal rational functions \\}
\label{Nonlin}
The series expansion (\ref{reduc7Aser}) of the 
ratio of two $\, _2F_1$ hypergeometric series of section \ref{compare6}
\begin{eqnarray}
\label{reduc7Aapp}
\hspace{-0.75in}&& \quad \quad \quad \quad \quad \quad
R(x) \, \, = \, \,  \,  \, \, 
{{ _2F_1\Bigl([{{1} \over {3}}, \,{{1} \over {3}}], \, [1], \, \, 27 \, x \Bigr) 
} \over {
 _2F_1\Bigl([{{1} \over {2}}, \,{{1} \over {2}}], \, [1], \, \, 16 \, x \Bigr)}}.
\end{eqnarray}
is {\em solution of the non-linear differential equation}
($\, R$ denotes $\,R(x)$, and $\, R_n$ 
denote $\, d^nR/dx^n$):
\begin{eqnarray}
\label{nonlin}
\hspace{-0.95in}&& \, \, 
-2\,{x}^{2} \cdot \, (27\,x-1 )\,   \, (16\,x-1 )\,  \cdot \,\Bigl(
(27\,x-1 )\,  \cdot  \, (16\,x-1 ) \cdot \,  R_1 \, 
- \, ( 72\,x+1 )\cdot \,  R \Bigr) \cdot \,  R_3
\nonumber \\ 
\hspace{-0.95in}&& \quad \quad 
 -2\, x \cdot \, 
\Bigl(3 \,x \cdot \, (16\,x-1) \,   \, (72\,x+1) \,   \, (27\,x-1) \cdot \,  R_1 
\nonumber  \\ 
\hspace{-0.95in}&& \quad \quad \quad \quad \quad \quad \quad \quad 
- \, (93312\,{x}^{3}-168\,{x}^{2}-297\,x+4) \cdot \,  R\Bigr) \cdot \,  R_2
 \nonumber \\ 
\hspace{-0.95in}&&  \quad \quad \quad \quad 
\, \, +2\,\cdot \, (29376\,{x}^{3}+5580\,{x}^{2}-221\,x+1) \cdot \,  R \cdot \, R_1 
\nonumber \\ 
\hspace{-0.95in}&& \quad \quad \quad \quad 
\, \, + 3\,{x}^{2} \cdot \, (27\,x-1)^{2} \, (16\,x-1)^{2} \cdot \,  {R_2}^{2}
 \\ 
\hspace{-0.95in}&& \quad \quad \quad \quad 
\, \,+ \, ( 16\,x-1 )\,   \, ( 1944\,{x}^{3}-1569\,{x}^{2}+58\,x-1 ) \cdot \,  {R_1}^{2}
\nonumber \\ 
\hspace{-0.95in}&&   \quad \quad \quad \quad 
\, \, + \, ( 144\,{x}^{2}-432\,x+1 )\cdot \,  {R}^{2} \, \,\, \, = \, \,\, \, \, 0. 
\nonumber
\end{eqnarray}

\end{document}